\documentclass[graybox]{svmult}

\usepackage{amsfonts}
\usepackage{color}
\usepackage{amsmath}
\usepackage{amstext}
\usepackage{graphicx}
\usepackage{amssymb}
\usepackage{esint}
\usepackage[all]{xy}
\usepackage{hyperref}

\usepackage{mathptmx}       
\usepackage{helvet}         
\usepackage{courier}        
\usepackage{type1cm}        
\usepackage{makeidx}         
\usepackage{graphicx}        
\usepackage{multicol}        
\usepackage[bottom]{footmisc}

\makeindex             

\begin{document}

\title*{Polynomial Structure of Topological String Partition Functions}
\author{Jie Zhou}
\institute{Jie Zhou \at Department of Mathematics, Harvard University, One Oxford Street, MA 02138, USA, \email{jiezhou@math.harvard.edu}
}
%
%
\maketitle

\abstract{We review the polynomial structure of the topological string partition functions as solutions to the holomorphic anomaly equations. We also explain the connection between the ring of propagators defined from special K\"ahler geometry and the ring of almost-holomorphic modular forms defined on modular curves.}


\section{Introduction}

The last few decades have witnessed many exciting developments and applications of mirror symmetry \cite{Greene:1996cy, Cox:1999, Hori:2003ic, Alim:2012gq}. One of the prominent applications is using mirror symmetry to predict Gromov-Witten invariants initiated by the celebrated work \cite{Candelas:1990rm}.

Consider a smooth Calabi-Yau (CY) 3-fold $\check{X}$ sitting in a family of Calabi-Yau 3-folds
$\check{\pi}:
\check{\mathcal{X}}\rightarrow \check{\mathcal{M}}$, where
$\check{\mathcal{M}}$ is the moduli space of complexified K\"ahler structures\footnote{Throughout this note, we shall simply call them K\"ahler structures by abuse of language.} of $\check{X}$ whose dimension is $h^{1,1}(\check{X})$. The generating function of genus $g$ Gromov-Witten invariants gives the following quantity defined on the moduli space
\begin{equation}\label{AmodelFg}
\check{F}_{g}(\check{t})=\sum_{\beta\in H_{2}(\check{X},\beta)}
\langle e^{\omega}
\rangle_{g,\beta}\,,
\end{equation}
where
\begin{eqnarray*}
\omega=\sum_{i=1}^{h^{1,1}(\check{X})} \check{t}^{i}\omega_{i}\,,\quad
\langle \omega_{i_{1}}\omega_{i_{2}}\cdots \omega_{i_{k}}
\rangle_{g,\beta}=\prod_{j=1}^{k}ev_{i_{j}}^{*}\omega_{i_{j}}\cap [\overline{\mathcal{M}_{g,k}(\check{X},\beta)}]^{\mathrm{vir}}\,.
\end{eqnarray*}
Here
\begin{itemize}
\item
$\{\omega_{i}\}_{i=1}^{h^{1,1}(\check{X})}$ are the generators for the K\"ahler cone in the moduli space $\check{\mathcal{M}}$ of K\"ahler structure of $\check{X}$;
\item
$\check{t}=\{\check{t}^{i}\}_{i=1}^{h^{1,1}(\check{X})}$ are local coordinates on the moduli space $\check{\mathcal{M}}$;
\item
$[\overline{\mathcal{M}_{g,k}(\check{X},\beta)}]^{\mathrm{vir}}$
is the virtual fundamental class of the moduli space $\mathcal{M}_{g,k}(\check{X},\beta)$ of stable maps of genus $g$ and class $\beta$ with $k$ markings;
\item
$ev_{i_{j}},j=1,2,\cdots k$ are the evaluation maps: $\mathcal{M}_{g,k}(\check{X},\beta)
\rightarrow \check{X}$.
\end{itemize}
The quantity defined in Eq.\,(\ref{AmodelFg}) is {\it{a priori}} only a formal series defined near the point $e^{\check{t}^{i}}=0, i=1,2,\cdots, h^{1,1}(\check{X})$. That it represents a well-defined function is due to the mirror symmetry
conjecture which we shall review below.
An alternative way to write the above generating function $F_{g}(\check{t})$ in which the Gromov-Witten invariants appear naturally is the following
\begin{equation}\label{FgGWcheck}
\check{F}_{g}(\check{t})=\sum_{\beta\in H_{2}(\check{X},\mathbb{Z})}
N_{g,\beta}(\check{X})e^{2\pi i\int_{\beta} \omega}\,.
\end{equation}
Note that in this formula $N_{g,\beta}(\check{X})$ is independent of $\check{t}$ but depends only on the generic member $\check{X}$ in the family, this results from the fact that the Gromov-Witten invariants are deformation invariant.

The mirror symmetry conjecture predicts that for the CY 3-fold family (A-model) $\check{\pi}:\check{\mathcal{X}}\rightarrow \check{\mathcal{M}}$, there exists another family (B-model) of CY 3-folds $\pi:\mathcal{X}\rightarrow \mathcal{M}$ satisfying the following properties:
\subsubsection*{Mirror symmetry conjecture}
\begin{itemize}
\item
The moduli space $\check{\mathcal{M}}$
of K\"ahler structures of $\check{X}$ is identified
with the moduli space $\mathcal{M}$ of complex structures of a smooth CY 3-fold $X$ called the mirror manifold. This implies in particular that
$h^{1,1}(\check{X})=\dim \check{\mathcal{M}}=\dim \mathcal{M}=h^{2,1}(X)$.
\item
There exist distinguished coordinates $\check{t}
=\{\check{t}^{i}\}_{i=1}^{h^{1,1}(\check{X})}$
and $t
=\{t^{i}\}_{i=1}^{h^{2,1}(X)}$ called canonical coordinates (see \cite{Bershadsky:1993cx}) so that
the map $\check{t}=t$ gives the identification
$\check{\mathcal{M}}\cong \mathcal{M}$. This map is called the mirror map. In practice, one first matches some distinguished singular points on the moduli spaces, for example, the large volume limit on $\check{\mathcal{M}}$ and
the large complex structure limit on $\mathcal{M}$ (see Section \ref{sectiongenuszero}), then one identifies neighborhoods of these singular points by matching
the canonical coordinates on the moduli spaces.
\item
For each genus $g$, there  is a function $F_{g}(t)$ defined on the moduli space $\mathcal{M}$ so that under the mirror map, it
is identical to $\check{F}_{g}(\check{t})$.
\item
Moreover, topological string theory tells that the more natural objects one should be looking at on both sides are some non-holomorphic functions
$\check{\mathcal{F}}^{(g)}(\check{t},\bar{\check{t}})$
and $\mathcal{F}^{(g)}(t,\bar{t})$
which are again identical under the mirror map.
These quantities are called topological string partition functions.
The ``holomorphic limits" \cite{Bershadsky:1993ta, Bershadsky:1993cx} of the normalized partition functions give rise to the quantities $\check{F}_{g}(\check{t})$ and $F_{g}(t)$, respectively.
\end{itemize}

Determining the function $\mathcal{F}^{(g)}(t,\bar{t})$ is usually much easier than computing the Gromov-Witten invariants appearing in the
generating function $\check{F}_{g}(\check{t})$, since the latter requires a careful study of the moduli space of stable maps which is in general very complicated (see e.g.,  \cite{Kontsevich:1994na, Givental:1997, Lian:1997}), while the former satisfies some
recursive differential equations called holomorphic anomaly
equations \cite{Bershadsky:1993cx} which are relatively easier to handle,
as will be discussed below. These differential equations and the
corresponding boundary conditions were derived from physics, but can be formulated purely in terms
of mathematical language.

The general idea of counting curves via mirror symmetry is as follows. First one takes the holomorphic anomaly equations with boundary conditions as the defining equations for the topological string partition functions $\mathcal{F}^{(g)}(t,\bar{t})$.
Then one tries to solve for them from the equations. After that one normalizes them and takes the holomorphic limit at the large complex structure limit to get $F_{g}(t)$. Finally one uses the mirror map $\check{t}=t$ which matches the large volume limit with the large complex structure limit to obtain $\check{F}_{g}(\check{t})$, and thus extract the Gromov-Witten invariants $N_{g,\beta}(\check{X})$ from Eq.\,(\ref{FgGWcheck}).
In this way, via mirror symmetry, counting
curves in the CY 3-fold $\check{X}$ is translated into
solving the holomorphic anomaly equations on the moduli space
$\mathcal{M}$.
Interested readers are referred to \cite{Bershadsky:1993ta, Bershadsky:1993cx} for details on this subject.\footnote{See also \cite{Marino:1998pg, Katz:1999xq, Klemm:1999gm, Klemm:2004km, Yamaguchi:2004bt, Klemm:2005pd, Huang:2006hq, Aganagic:2006wq, Huang:2006si, Alim:2007qj, Grimm:2007tm, Alim:2008kp, Haghighat:2008gw, Haghighat:2009nr, Sakai:2011xg, Alim:2012ss, Klemm:2012sx, Alim:2013eja} for related works.}

This note aims to study some properties of the holomorphic anomaly equations and the solutions.
The plan of this note is as follows. In Section \ref{sectionHAE}, we review the basics of special K\"ahler geometry and holomorphic anomaly equations.
In Section \ref{sectionpolytech}, we explain the polynomial recursion technique and show how to solve the topological string partition functions genus by genus recursively from the holomorphic anomaly equations. In Section \ref{sectionmodularity}, we discuss the similarity between the ring of propagators and the ring of almost-holomorphic modular forms. We conclude this note in Section \ref{sectionconclusion}.
\medskip

This introductory note is based on the lectures that the author gave in the Concentrated
Graduate Courses for the Fields thematic program
Calabi-Yau Varieties: Arithmetic, Geometry and
Physics at the Fields Institute in Toronto.
None of the material presented in this note is original and the author apologizes in advance for everything that may have been left out or not been attributed correctly.

\medskip
\textbf{Acknowledgements} The author would like to thank Murad Alim, Emanuel
Scheidegger and Shing-Tung Yau for valuable collaborations and
inspiring discussions on related projects.
Thanks also goes to Murad Alim, Emanuel Scheidegger, Teng Fei and Atsushi Kanazawa for carefully reading the draft and giving very helpful comments.
He also wants to thank Professor Noriko Yui and the other organizers for inviting him to the thematic program
Calabi-Yau Varieties: Arithmetic, Geometry and
Physics at the Fields Institute, and
the Fields Institute for providing an excellent research atmosphere and partial financial support during his visiting.


\section{Holomorphic anomaly equations}
\label{sectionHAE}

In this section, we shall first review briefly the special K\"ahler geometry on the deformation space of CY 3-folds and the basics of classical genus zero mirror symmetry.
After that we shall display the holomorphic anomaly equations satisfied by the topological string partition functions.

\subsection{Special K\"ahler geometry}

Consider a family $\pi:\mathcal{X}\rightarrow \mathcal{M}$ of
CY 3-folds $\mathcal{X}=\{\mathcal{X}_{z}\}$ over a
variety $\mathcal{M}$ parametrized by the complex coordinate system
$z=\{z^{i}\}_{i=1}^{\dim \mathcal{M}}$.
For a generic $z\in
\mathcal{M}$, the fiber $\mathcal{X}_{z}$ is a smooth CY
3-fold.
We also assume that $\dim
\mathcal{M}=h^{1}(\mathcal{X}_{z},T\mathcal{X}_{z})$ for a smooth
$\mathcal{X}_{z}$, where $T\mathcal{X}_{z}$ is the holomorphic
tangent bundle of $\mathcal{X}_{z}$. In the following, we shall use the
notation $X$ to denote a generic fiber $\mathcal{X}_{z}$ in the
family without specifying the point $z$.

In the examples discussed in this note, the smooth CY 3-fold $X$ is toric in nature, i.e., it is a hypersurface or complete intersection in a toric variety, and
the variety $\mathcal{M}$ will be the moduli space of complex structure of $X$ which can be constructed torically.

The variation of complex structure on $X$ can be studied by looking at the periods according to the general theory of variation of Hodge structures.
They are defined to be
the integrals $\Pi=\int_{C}\Omega_{z}$, where $C\in
H_{3}(\mathcal{X}_{z},\mathbb{Z})$ and $\Omega=\{\Omega_{z}\}$ is a
holomorphic section of the Hodge line bundle
$\mathcal{L}=\mathcal{R}^{0}\pi_{*}\Omega^{3}_{\mathcal{X}|\mathcal{M}}$
on $\mathcal{M}$. They satisfy a differential equation system  $\mathcal{L}_{\textrm{CY}}\,\Pi=0$ called the
Picard-Fuchs equations induced from the Gauss-Manin connection on the
Hodge bundle
\begin{equation}
\mathcal{H}=\mathcal{R}^{3}\pi_{*}\underline{\mathbb{C}}\otimes
\mathcal{O}_{\mathcal{M}}
=
\mathcal{R}^{3}\pi_{*}\Omega_{\mathcal{X}|\mathcal{M}}^{\bullet}=\mathcal{L}\oplus \mathcal{L}\otimes T\mathcal{M}
\oplus  \overline{\mathcal{L}\otimes T\mathcal{M}}\oplus \overline{\mathcal{L}}\,.
\end{equation}
The base $\mathcal{M}$ of the family is equipped with the Weil-Petersson metric
whose K\"ahler potential $K$ is determined from
\begin{equation}\label{Kahlerpotential}
e^{-K(z,\bar{z})}=i\int_{\mathcal{X}_{z}} \Omega_{z}\wedge
\overline{\Omega}_{z}\,,
\end{equation}
where as above $\Omega=\{\Omega_{z}\}$ is a section of the Hodge line bundle
$\mathcal{L}$.
 The metric
$G_{i\bar{j}}=\partial_{i}\bar{\partial}_{\bar{j}}K$ is
the Hodge metric induced from the Hermitian metric
$h(\Omega,\Omega)=i^{3}\int \Omega\wedge \overline{\Omega}$ on the
Hodge line bundle $\mathcal{L}$. This metric is called special
K\"ahler metric \cite{Strominger:1990pd, Freed:1999sm}. Among its
other properties, it satisfies the following ``special geometry
relation''
\begin{equation}\label{specialgeometryrelation}
-R_{i\bar{j}~l}^{~~k}=\partial_{\bar{j}}\Gamma_{il}^{k}=\delta^{k}_{l}G_{i\bar{j}}+\delta^{k}_{i}G_{l\bar{j}}
-C_{ilm}\bar{C}_{\bar{j}}^{mk},\quad
i,\bar{j},k,l=1,2,\cdots \dim \mathcal{M}\,,
\end{equation}
where
\begin{equation}\label{dfnofYukawa}
C_{ijk}(z)=-\int_{\mathcal{X}_{z}}\Omega_{z}\wedge
\partial_{i}\partial_{j}\partial_{k}\Omega_{z}
\end{equation}
is the so-called Yukawa coupling and
\begin{equation}
\bar{C}_{\bar{j}}^{mk}=e^{2K}G^{k\bar{k}}G^{m\bar{m}}\bar{C}_{\bar{j}\bar{k}\bar{m}}.
\end{equation}
 Note that $C_{ijk}\in \Gamma(\mathcal{M},  \mathrm{Sym}^{\otimes 3}T^{*}\mathcal{M} \otimes\mathcal{L}^{2})$ and it is symmetric in $i,j,k$  by definition.
Integrating Eq.\,(\ref{specialgeometryrelation}), one then gets the ``integrated special geometry relation''
\begin{equation}\label{integratedspecialgeometryrelation}
\Gamma_{ij}^{k}=\delta_{j}^{k}K_{i}+\delta_{i}^{k}K_{j}-C_{ijm}S^{mk}+s_{ij}^{k}\,,
\end{equation}
where $S^{mk}$ is defined to be a solution to $\bar{\partial}_{\bar{n}}S^{mk}=\bar{C}_{\bar{n}}^{mk}$, and $s_{ij}^{k}$ could be any holomorphic quantity.
There is a natural covariant derivative $D$ acting on sections
of the Hodge bundle
$\mathcal{H}=\mathcal{R}^{3}\pi_{*}\underline{\mathbb{C}}\otimes
\mathcal{O}_{\mathcal{M}}$: it is induced from the Chern connection
associated to the Weil-Petersson metric and the connection on
$\mathcal{L}$ induced by the Hermitian metric $h=e^{-K}$.
For example, on a section $T^{I}_{J}$ of $\mathrm{Sym}^{\otimes k}T\mathcal{M}  \otimes\mathrm{Sym}^{\otimes l}T^{*}\mathcal{M}  \otimes  \mathcal{L}^{m}
 \otimes  \bar{\mathcal{L}}^{n}$, where $I=\{i_{1},i_{2},\cdots i_{k}\}, J=\{j_{1},i_{2},\cdots j_{l}\}$, one has
\begin{eqnarray*}
D_{i}T^{I}_{J}&=&\partial_{i}T^{I}_{J}+\sum_{i_{r}\in I}  \Gamma_{i a}^{i_{r}}    T^{a i_{1}i_{2}\cdots \hat{i_{r}}\cdots  i_{k} }_{J}
-
\sum_{j_{s}\in J}  \Gamma_{i b}^{j_{s}}    T^{b j_{1}j_{2}\cdots \hat{j_{s}}\cdots  j_{l} }_{J}
+
mK_{i} T^{I}_{J}\,,\\
\bar{D}_{\bar{i}}T^{I}_{J}&=&\bar{\partial}_{\bar{i}}T^{I}_{J}+
nK_{\bar{i}} T^{I}_{J}\,,
\end{eqnarray*}
where $\hat{}~$ means the index is excluded.
We have similar formulas for tensors with anti-holomorphic indices.

Eq.\,(\ref{specialgeometryrelation}) implies that there exists a holomorphic quantity $\mathcal{F}\in\Gamma(\mathcal{M},  \mathcal{L}^{2})$ called prepotential such that
\begin{equation}
C_{ijk}=D_{i}D_{j}D_{k}\mathcal{F}.
\end{equation}
See
\cite{Strominger:1990pd, Freed:1999sm, Hosono:2008ve} for details on
this.

\subsection{Variation of Hodge structures and genus zero mirror symmetry}\label{sectiongenuszero}

Among the singular points on the moduli space $\mathcal{M}$, there
is a distinguished point called the large complex structure limit.
It is mirror to the large volume limit (in terms of the coordinates $\{\check{t}^{i}\}_{i=1}^{h^{1,1}(\check{X})}$ in Eq.\,(\ref{AmodelFg}), the point is given by setting all
$\check{t}^{i}$s to infinity) on the mirror side (A-side)
and is a maximally unipotent monodromy point \cite{Morrison:1992}. It plays a special role in genus zero mirror symmetry.
The asymptotic behavior of the period map near this point
can be studied using the theory of variation of Hodge structures, see \cite{Schmid:1973} for the mathematical foundation and
\cite{Candelas:1990rm} (also \cite{Cox:1999, Gross:2003} for nice reviews) for applications of this in mirror symmetry.

Assume the large complex structure limit is given by $z=0$. Near this
point, the solutions to the Picard-Fuchs system $\mathcal{L}_{\textrm{CY}}\,
\Pi=0$ could be obtained by the Frobenius method and have the
following form
\begin{equation}\label{periodsforcompactCYs}
(X^{0},X^{a},P_{a},P_{0})=X^{0}(1,t^{a},\partial_{t^{a}}F(t),2F-t^{a}\partial_{t^{a}}F(t)), \quad a=1,2\cdots \dim \mathcal{M}\,,
\end{equation}
where $X^{0}(z)\sim 1+\mathcal{O}(z)$, while $t^{a} \sim (1 /2\pi i)\log
z^{a}+\textrm{regular}$ near $z=0$ gives local coordinates on the
punctured moduli space $\mathcal{M}-\{z=0\}$.
The existence of the holomorphic
function $F(t)$, called prepotential, and the above particular structure
of the periods
result from the special
K\"ahler geometry on $\mathcal{M}$.
The coordinates $t=\{t^{a}\}_{a=1}^{\dim \mathcal{M}}$ are in fact the canonical coordinates based at the large complex structure limit, see \cite{Bershadsky:1993ta, Bershadsky:1993cx} for details on this.
The prepotential $F$ here is related to the previously defined quantity $\mathcal{F}$ by
$F=X_{0}^{-2}\mathcal{F}\in \Gamma(\mathcal{M},\mathcal{L}^{0})$.

One can easily solve for the Yukawa couplings $C_{ijk}(z)$ from the Picard-Fuchs equation, see e.g., \cite{Cox:1999, Gross:2003}.
The normalized (so that after normalization it gives a section of $\mathcal{L}^{0}$) Yukawa coupling in the $t$ coordinates is then given by
\begin{equation}
C_{t^{a}t^{b}t^{c}}=\partial^{3}_{t^{a}t^{b}t^{c}}F=\frac{1}{ (X^{0})^{2}}C_{z^{i}z^{j}z^{k}}\frac{\partial z^{i}}{ \partial t^{a}}
\frac{\partial z^{j}}{ \partial t^{b}}
\frac{\partial z^{k}}{\partial t^{c}}\in \Gamma(\mathcal{M},\mathcal{L}^{0})\,.
\end{equation}
By equating the normalized Yukawa couplings from both the A-side and the B-side,
one can establish genus zero mirror symmetry under the mirror map $\check{t}=t$:
\begin{equation}
C_{t^{a}t^{b}t^{c}}=
\kappa_{abc}+\sum_{\beta\in H_{2}(\check{X},\mathbb{Z})}^{\infty}d_{a}d_{b}d_{c}N_{0,d_{a}d_{b}d_{c}}e^{2\pi i d_{a}t^{a}}\,,
\end{equation}
where $\kappa_{abc}=\omega_{a}\cup \omega_{b}\cup \omega_{c}$ is the classical triple intersection number of the
mirror manifold $\check{X}$ of $X$, $d_{a}=\int_{\beta}\omega_{a}$,
and $N_{0, d_{a}d_{b}d_{c}}$ is the same as the quantity $N_{0,\beta}$ in Eq.\,(\ref{FgGWcheck}).
This prediction has been checked for many CY 3-fold families by directly computing $N_{0,\beta}$ using techniques from the A-side, e.g., the localization technique \cite{Kontsevich:1994na}.

\subsection{Holomorphic anomaly equations}
According to \cite{Bershadsky:1993ta, Bershadsky:1993cx},
the genus $g$ topological string partition function\footnote{The quantity $\mathcal{F}^{(g)}$ is really a section rather than a function,
but in the literature it is termed topological string partition function
which we shall follow in this note.}
$\mathcal{F}^{(g)}$is a (smooth) section of the line bundle
$\mathcal{L}^{2-2g}$ over $\mathcal M$, it is shown to satisfy the
following holomorphic anomaly equation
\footnote{In this note, we shall use $\bar{\partial}_{\bar{\imath}}$ and $\partial_{\bar{\imath}}$ interchangeably to denote
$\frac{\partial}{\partial \bar{z}^{\bar{\imath}}}$ for some local complex coordinates $z=\{z^{i}\}_{i=1}^{\mathrm{dim} \mathcal{M}}$  chosen on the moduli space $\mathcal{M}$. }:
\begin{eqnarray}
\bar{\partial}_{\bar{\imath}} \partial_{j}\mathcal{F}^{(1)} &=&
\frac{1}{2} C_{jkl} \overline{C}^{kl}_{\bar{\imath}}+
(1-\frac{\chi}{24}) G_{j
\bar{\imath}}\,,\label{haeforgenusone}\\
\bar{\partial}_{\bar{\imath}} \mathcal{F}^{(g)} &=&{1\over 2}
\overline{C}_{\bar{\imath}}^{jk} \left( \sum_{r=1}^{g-1}
D_j\mathcal{F}^{(r)}\,  D_k\mathcal{F}^{(g-r)} +
D_j D_k\mathcal{F}^{(g-1)} \right), \quad g\geq 2\,,\label{haeforhighergenus}
\end{eqnarray}
where $\chi$ is the Euler characteristic of the mirror manifold $\check{X}$ of
the CY 3-fold $X$.

As we shall see below, any genus $\mathcal{F}^{(g)}$
can be determined recursively from these equations up to addition by a
holomorphic function $f^{(g)}$ called holomorphic ambiguity.
Boundary conditions on the (global) moduli space are needed to fix the holomorphic ambiguity
$f^{(g)}$. What are commonly used are the asymptotic behaviors of $\mathcal{F}^{(g)}$ at the
singular points on the moduli space $\mathcal{M}$, see \cite{Bershadsky:1993ta, Bershadsky:1993cx, Ghoshal:1995wm, Huang:2006si, Huang:2006hq}.

\section{Polynomial structure of topological string partition functions}
\label{sectionpolytech}
We now shall explain the polynomial recursion technique which was developed in \cite{Yamaguchi:2004bt, Alim:2007qj} to solve the holomorphic anomaly equations in Eqs.\,(\ref{haeforgenusone}, \ref{haeforhighergenus}).

\subsection{Propagators}
First let us try to solve for genus one and two topological string partition function from the holomorphic anomaly equations,
we shall see in the sequel why it is convenient to introduce the so-called propagators.

Consider the genus one topological string partition function, in order to solve
\begin{equation*}
\bar{\partial}_{\bar{i}}\partial_{j} \mathcal{F}^{(1)} =
\frac{1}{2} C_{jkl} \overline{C}^{kl}_{\bar{i}}+
(1-\frac{\chi}{24}) G_{j\bar{i}}\,,
\end{equation*}
one needs to turn the right hand side of the equation into a total derivative (with respect to $\bar{\partial}_{\bar{i}}$).
This then leads to the following definition of the propagator $S^{kl}$:
\begin{equation}\label{defofSij}
\partial_{\bar{i}}S^{kl}=\overline{C}^{kl}_{\bar{i}}.
\end{equation}
Note that this implies in particular that $S^{kl}$ is a (non-holomorphic) section of $\mathcal{L}^{-2}\otimes \mathrm{Sym}^{2}T\mathcal{M}$. Then we get
\begin{equation}\label{soltogenusone}
\bar{\partial}_{\bar{i}}\partial_{j}\mathcal{F}^{(1)}=\bar{\partial}_{\bar{i}}
\left({1\over 2}C_{jkl}S^{kl}+(1-{\chi\over 24})K_{j}\right)\,.
\end{equation}
From this one can see that there exists some holomorphic quantity $f_{j}^{(1)}$ called genus one holomorphic ambiguity so that
\begin{equation}\label{integratedgenusone}
\partial_{j}\mathcal{F}^{(1)}=
{1\over 2}C_{jkl}S^{kl}+(1-{\chi\over 24})K_{j} +f_{j}^{(1)}.
\end{equation}
The above genus one result is proved rigorously in mathematics for a large class of Calabi-Yau manifolds, see \cite{Fang:2008, Zinger:2008, Zinger:2009, Popa:2013} for some progress.

Now let us proceed to genus two holomorphic anomaly equation. Recall that $\mathcal{F}^{(1)}$
is a section of $\mathcal{L}^{0}$, we know $D_{j}\mathcal{F}^{(1)}=\partial_{j}\mathcal{F}^{(1)}$. Hence
the holomorphic anomaly equation for genus two simplifies to
\begin{eqnarray*}
\bar{\partial}_{\bar{i}} \mathcal{F}^{(2)} &=&{1\over 2}
\overline{C}_{\bar{i}}^{jk} \left(
\partial_j\mathcal{F}^{(1)} \partial_k\mathcal{F}^{(1)} +
D_j \partial_k\mathcal{F}^{(1)} \right)\\
&=&{1\over 2}
\overline{C}_{\bar{i}}^{jk} \left(
\partial_j\mathcal{F}^{(1)} \partial_k\mathcal{F}^{(1)} +
\partial_j \partial_k\mathcal{F}^{(1)} -\Gamma_{jk}^{l}\partial_l\mathcal{F}^{(1)}\right).
\end{eqnarray*}
Again one needs to turn
the right hand side of the equation into a total derivative. Note that $S^{jk}$ is symmetric in $j,k$,
using Eq.\,(\ref{defofSij}) and integration by parts, we then get
\begin{eqnarray*}
\bar{\partial}_{\bar{i}} \mathcal{F}^{(2)}
&=&
{1\over 2}\partial_{\bar{i}}    \left(  S^{jk}(\partial_j\mathcal{F}^{(1)} \partial_k\mathcal{F}^{(1)}
+
\partial_j \partial_k\mathcal{F}^{(1)} -\Gamma_{jk}^{l}\partial_l\mathcal{F}^{(1)})
\right)\\
&+&
\underbrace{
- S^{jk}  \partial_{\bar{i}}
\partial_j\mathcal{F}^{(1)} \partial_k\mathcal{F}^{(1)}
}_{\mathrm{I}} +
\underbrace{
-{1\over 2} S^{jk} \partial_{\bar{i}}  \partial_j \partial_k\mathcal{F}^{(1)}
}_{\mathrm{II}} \\
&+&
\underbrace{
{1\over 2} S^{jk}\partial_{\bar{i}} \Gamma_{jk}^{l}\partial_l\mathcal{F}^{(1)}
}_{\mathrm{III}}
+
\underbrace{
{1\over 2} S^{jk}\Gamma_{jk}^{l}\partial_{\bar{i}}  \partial_l\mathcal{F}^{(1)}
}_{\mathrm{IV}} .
\end{eqnarray*}
Then we can plug in the special geometry relation Eq.\,(\ref{specialgeometryrelation}), the integrated special geometry relation Eq.\,(\ref{integratedspecialgeometryrelation}), the holomorphic anomaly equation for
$\mathcal{F}^{(1)}$ given in Eq.\,(\ref{haeforgenusone}) and Eq.\,(\ref{integratedgenusone}) to simplify the above expressions $\mathrm{I}-\mathrm{IV}$.
For simplicity, we shall denote $c_{\chi}:=1-{\chi\over 24}$.
The terms $\mathrm{I}, \mathrm{II}, \mathrm{III},\mathrm{IV}$ are given by
\begin{eqnarray*}
\mathrm{I}
&=&
 -S^{jk}
( {1\over 2}
C_{jmn} \partial_{\bar{i}} S^{mn} +c_{\chi} \partial_{\bar{i}}K_{j})
({1\over 2}  C_{krs}S^{rs}+c_{\chi}K_{k}+f_{k}^{(1)})\\
&=&
 -{1\over 4}C_{jmn}C_{krs}\partial_{\bar{i}} S^{mn}   S^{jk} S^{rs}
-{1\over 2}c_{\chi} C_{jmn}  \partial_{\bar{i}} S^{mn}     S^{jk} K_{k}
-{1\over 2} f_{k}^{(1)} C_{jmn}  \partial_{\bar{i}} S^{mn}     S^{jk}  \\
&&
-{1\over 2} c_{\chi} C_{krs}  \partial_{\bar{i}} K_{j}S^{jk} S^{rs}
- c_{\chi}^{2}  \partial_{\bar{i}} K_{j}S^{jk} K_{k}
-c_{\chi}   f_{k}^{(1)} S^{jk} \partial_{\bar{i}} K_{j}\\
\mathrm{II}&=&
-{1\over 2}S^{jk}  \partial_{k}  (   {1\over 2}C_{jmn} \partial_{\bar{i}} S^{mn}+c_{\chi} \partial_{\bar{i}}K_{j} )\,,\\
\mathrm{III}&=&
{1\over 2}S^{jk}  (\delta_{j}^{l}\partial_{\bar{i}} K_{k}+\delta_{k}^{l}\partial_{\bar{i}} K_{j} -C_{jkp}\partial_{\bar{i}}S^{pl})
({1\over 2} C_{lmn}S^{mn}+c_{\chi} K_{l}+f_{l}^{(1)})\,,\\
\mathrm{IV}&=&
{1\over 2}S^{jk} (\delta_{j}^{l} K_{k}+\delta_{k}^{l} K_{j} -C_{jkp}S^{pl}+s_{jk}^{l}) ({1\over 2}C_{lmn}\partial_{\bar{i}} S^{mn}+c_{\chi}\partial_{\bar{i}}  K_{l} )\,.
\end{eqnarray*}
Now we think of all the terms as formal polynomials in $S^{jk}, K_{j}$ and their holomorphic and anti-holomorphic derivatives.
In the sum $\mathrm{I}+\mathrm{II}+\mathrm{III}+\mathrm{IV}$ there is no monomial of the form $\partial_{\bar{i}}S^{jk}  K_{j}$.
Moreover, one can show that the coefficient of the monomial $S^{jk}\partial_{\bar{i}}  K_{j}$ which involves $f_{k}^{(1)},s_{jk}^{j}$ is not vanishing for general geometries.
It follows that terms of the form $S^{jk}\partial_{\bar{i}}  K_{j} $ must be total derivatives with respect
to $\partial_{\bar{i}}$. This leads to the following definition of the propagator $S^{j}$ which is a non-holomorphic section of $\mathcal{L}^{-2}\otimes T\mathcal{M}$.:
\begin{equation}\label{defofSi}
\partial_{\bar{i}} S^{j}=G_{\bar{i}k}S^{jk}\,.
\end{equation}
Then we can further simplify the quantities $\mathrm{I}-\mathrm{IV}$ as follows:
\begin{eqnarray*}
\mathrm{I}
&=&
 -{1\over 4}C_{jmn}C_{krs}\partial_{\bar{i}} S^{mn}   S^{jk} S^{rs}
-{1\over 2}c_{\chi} C_{jmn}  \partial_{\bar{i}} S^{mn}     S^{jk} K_{k}
-{1\over 2} f_{k}^{(1)} C_{jmn}  \partial_{\bar{i}} S^{mn}     S^{jk}  \\
&&
-{1\over 2} c_{\chi} C_{krs}  \partial_{\bar{i}} K_{j} S^{jk}S^{rs}
- c_{\chi}^{2}  \partial_{\bar{i}} S^{j} K_{j}
-c_{\chi}   f_{k}^{(1)} \partial_{\bar{i}} S^{k}\,,\\
\mathrm{II}&=&
-{1\over 4}S^{jk}  \partial_{k}\partial_{\bar{i}}  ( C_{jmn}  S^{mn})-{1\over 2}c_{\chi} S^{jk}\partial_{k}\partial_{\bar{i}}K_{j} \,,\\
\mathrm{III}&=&{1\over 2}
\partial_{\bar{i}} K_{k} S^{jk}  C_{jmn}S^{mn}+c_{\chi}K_{j}\partial_{\bar{i}} S^{j}+f_{j}^{(1)}\partial_{\bar{i}} S^{j}\\
&&
-{1\over 4}C_{jkp}C_{lmn}S^{jk}S^{mn}\partial_{\bar{i}} S^{pl}
-{1\over 2}c_{\chi}C_{jkp}S^{jk}\partial_{\bar{i}} S^{pl} K_{l}
-{1\over 2}C_{jkp}f_{l}^{(1)}S^{jk}\partial_{\bar{i}}  S^{pl}
\,,\\
\mathrm{IV}&=&
{1\over 2}C_{jmn}S^{jk}  K_{k}\partial_{\bar{i}} S^{mn}
-{1\over 4} C_{lmn}C_{jkp}S^{jk}S^{pl}\partial_{\bar{i}} S^{mn}
+{1\over 4} C_{lmn}s_{jk}^{l}S^{jk}\partial_{\bar{i}} S^{mn}\\
&&
+c_{\chi} K_{j}\partial_{\bar{i}} S^{j}
-{1\over 2} c_{\chi} C_{jkp}S^{jk}\partial_{\bar{i}}  K_{l} S^{pl}+{1\over 2}c_{\chi}S^{jk}s_{jk}^{l}\partial_{\bar{i}}  K_{l}\,.
\end{eqnarray*}
Again to turn $K_{j}\partial_{\bar{i}} S^{j}=\partial_{\bar{i}} (K_{j}S^{j})-S^{j}\partial_{\bar{i}} K_{j}$ into a total derivative,
we need to define a propagator $S$ which is a non-holomorphic section of $\mathcal{L}^{-2}$ so that
\begin{equation}\label{defofS}
\partial_{\bar{i}} S=G_{\bar{i}j}S^{j}\,.
\end{equation}
Now the terms $\mathrm{I}, \mathrm{II}$ become
\begin{eqnarray*}
\widetilde{\mathrm{I}}
&=&
 -{1\over 4}C_{jmn}C_{krs}\partial_{\bar{i}} S^{mn}   S^{jk} S^{rs}
-{1\over 2}c_{\chi} C_{jmn}  \partial_{\bar{i}} S^{mn}     S^{jk} K_{k}
-{1\over 2} f_{k}^{(1)} C_{jmn}  \partial_{\bar{i}} S^{mn}     S^{jk}  \\
&&
-{1\over 2} c_{\chi} C_{krs}  \partial_{\bar{i}} K_{j}   S^{jk}S^{rs}
- c_{\chi}^{2}  \partial_{\bar{i}} (S^{k}K_{k})+ c_{\chi}^{2}  \partial_{\bar{i}} S
-c_{\chi}    \partial_{\bar{i}} (f_{k}^{(1)}S^{k})\,,\\
\widetilde{\mathrm{II}}&=&
-{1\over 4}S^{jk}  \partial_{k}\partial_{\bar{i}}  ( C_{jmn}  S^{mn})-{1\over 2}c_{\chi} S^{jk}\partial_{k}\partial_{\bar{i}}K_{j} \,.
\end{eqnarray*}

It turns out that after introducing these propagators, the quantity
$\widetilde{\mathrm{I}}+\widetilde{\mathrm{II}}+\mathrm{III}+\mathrm{IV}$ becomes a total derivative. Before we show this, we make a pause and study the derivatives of
quantities $S^{jk},S^{j},S,K_{i}$. They will be needed later, for example, to simplify $\widetilde{\mathrm{II}}$.

\subsection{Differential ring of propagators}

We use the definition of the propagators Eqs.\,(\ref{defofSij}, \ref{defofSi}, \ref{defofS}) and the relations Eqs.\,(\ref{specialgeometryrelation}, \ref{integratedspecialgeometryrelation}) to derive the differential ring structure.
The results in this section follow the presentation in
\cite{Alim:2007qj}.

To obtain an expression for $D_{i}S^{jk}$ in terms of known quantities, we compute $\bar{\partial}_{\bar{l}}D_{i}S^{jk}$ by commuting the derivatives, then we
integrate what we get to obtain the desired expression.
Since by definition $S^{jk}$ is a section of $\mathcal{L}^{-2}\otimes \mathrm{Sym}^{\otimes 2}T\mathcal{M}$, we know when acting on $S^{jk}$ the covariant derivative is given by $D_{i}=\partial_{i}+\Gamma_{i}+(-2)K_{i}$,
hence
\begin{equation*}
\bar{\partial}_{\bar{l}}D_{i}S^{jk}
=\bar{\partial}_{\bar{l}}(\partial_{i}S^{jk}+\Gamma^{j}_{im}S^{mk}+\Gamma^{k}_{in}S^{jn}-2K_{i}S^{jk})
:=A+B+C+D\,.
\end{equation*}
Then
\begin{eqnarray*}
A&=&\partial_{i}\bar{C}^{jk}_{\bar{l}}=\partial_{i}(e^{2K}G^{j\bar{j}}G^{k\bar{k}}\bar{C}_{\bar{j}\bar{k}\bar{l}})\\
&=&2K_{i}\bar{C}^{jk}_{\bar{l}}+e^{2K}\partial_{i}G^{j\bar{j}}G^{k\bar{k}}\bar{C}_{\bar{j}\bar{k}\bar{l}}+e^{2K}G^{j\bar{j}}
\partial_{i}G^{k\bar{k}}\bar{C}_{\bar{j}\bar{k}\bar{l}}+e^{2K}G^{j\bar{j}}G^{k\bar{k}}\partial_{i}\bar{C}_{\bar{j}\bar{k}\bar{l}}\,.
\end{eqnarray*}
Since $C_{ijk}$ is holomorphic, $\bar{C}_{\bar{j}\bar{k}\bar{l}}$ is anti-holomorphic, the last term above is $0$.
It follows that
\begin{eqnarray*}
A&=&2K_{i}\bar{C}^{jk}_{\bar{l}}+e^{2K}\partial_{i}G^{j\bar{j}}G^{k\bar{k}}\bar{C}_{\bar{j}\bar{k}\bar{l}}+e^{2K}G^{j\bar{j}}
\partial_{i}G^{k\bar{k}}\bar{C}_{\bar{j}\bar{k}\bar{l}}+e^{2K}G^{j\bar{j}}G^{k\bar{k}}\partial_{i}\bar{C}_{\bar{j}\bar{k}\bar{l}}\\
&=&2K_{i}\bar{C}^{jk}_{\bar{l}}+e^{2K}(-\Gamma^{j}_{ir}G^{r\bar{j}})G^{k\bar{k}}\bar{C}_{\bar{j}\bar{k}\bar{l}}+e^{2K}G^{j\bar{j}}
(-\Gamma^{k}_{is}G^{s\bar{k}})\bar{C}_{\bar{j}\bar{k}\bar{l}}+0\,.
\end{eqnarray*}
Similarly,
\begin{eqnarray*}
B&=&\bar{\partial}_{\bar{l}}\Gamma^{j}_{im}S^{mk}+\Gamma^{j}_{im}\bar{\partial}_{\bar{l}}S^{mk}=(\delta^{j}_{i}G_{m\bar{l}}+\delta^{j}_{m}G_{i\bar{l}}-C_{imt}\bar{C}^{tj}_{\bar{l}})S^{mk}+\Gamma^{j}_{im}\bar{C}^{mk}_{\bar{l}}\,,\\
C&=&\bar{\partial}_{\bar{l}}\Gamma^{k}_{in}S^{nj}+\Gamma^{k}_{in}\bar{\partial}_{\bar{l}}S^{nj}=(\delta^{k}_{i}G_{n\bar{l}}+\delta^{k}_{n}G_{i\bar{l}}-C_{int}\bar{C}^{tk}_{\bar{l}})S^{nj}+\Gamma^{k}_{in}\bar{C}^{nj}_{\bar{l}}\,,\\
D&=&-2G_{i\bar{l}}S^{jk}-2K_{i}\bar{C}^{jk}_{\bar{l}}\,.
\end{eqnarray*}
It follows that
\begin{eqnarray*}
&&A+B+C+D\\
&=&K_{i}\bar{C}^{jk}_{\bar{l}}+e^{2K}(-\Gamma^{j}_{ir}G^{r\bar{j}})G^{k\bar{k}}\bar{C}_{\bar{j}\bar{k}\bar{l}}+e^{2K}G^{j\bar{j}}
(-\Gamma^{k}_{is}G^{s\bar{k}})\bar{C}_{\bar{j}\bar{k}\bar{l}}+0\\
&&+\delta^{j}_{i}G_{m\bar{l}}S^{mk}+\delta^{j}_{m}G_{i\bar{l}}S^{mk}-C_{imt}\bar{C}^{tj}_{\bar{l}}S^{mk}+\Gamma^{j}_{im}\bar{C}^{mk}_{\bar{l}}\\
&&+\delta^{k}_{i}G_{n\bar{l}}S^{nj}+\delta^{k}_{n}G_{i\bar{l}}S^{nj}-C_{int}\bar{C}^{tk}_{\bar{l}}S^{nj}+\Gamma^{k}_{in}\bar{C}^{nj}_{\bar{l}}\quad -2G_{i\bar{l}}S^{jk}-2K_{i}\bar{C}^{jk}_{\bar{l}}\\
&=&{K_{i}\bar{C}^{jk}_{\bar{l}}}+e^{2K}(-\Gamma^{j}_{ir}G^{r\bar{j}})G^{k\bar{k}}\bar{C}_{\bar{j}\bar{k}\bar{l}}+e^{2K}G^{j\bar{j}}
(-\Gamma^{k}_{is}G^{s\bar{k}})\bar{C}_{\bar{j}\bar{k}\bar{l}}+0\\
&&+\delta^{j}_{i}G_{m\bar{l}}S^{mk}+{\delta^{j}_{m}G_{i\bar{l}}S^{mk}}-C_{imt}\bar{C}^{tj}_{\bar{l}}S^{mk}+\Gamma^{j}_{im}\bar{C}^{mk}_{\bar{l}}\\
&&+\delta^{k}_{i}G_{n\bar{l}}S^{nj}+{\delta^{k}_{n}G_{i\bar{l}}S^{nj}}-C_{int}\bar{C}^{tk}_{\bar{l}}S^{nj}+\Gamma^{ki}_{in}\bar{C}^{nj}_{\bar{l}}
\quad {-2G_{i\bar{l}}S^{jk}}{-2K_{i}\bar{C}^{jk}_{\bar{l}}}\,.
\end{eqnarray*}
Using the definition
$G_{m\bar{l}}S^{mk}=S^{k}_{\bar{l}}=\bar{\partial}_{\bar{l}}S^{k}$, we get
\begin{eqnarray*}
&&A+B+C+D\\
&=&{K_{i}\bar{C}^{jk}_{\bar{l}}}+e^{2K}(-\Gamma^{j}_{ir}G^{r\bar{j}})G^{k\bar{k}}\bar{C}_{\bar{j}\bar{k}\bar{l}}+e^{2K}G^{j\bar{j}}
(-\Gamma^{k}_{is}G^{s\bar{k}})\bar{C}_{\bar{j}\bar{k}\bar{l}}+0\\
&&+\delta^{j}_{i}\bar{\partial}_{\bar{l}}S^{k}+{\delta^{j}_{m}G_{i\bar{l}}S^{mk}}-C_{imt}\bar{C}^{tj}_{\bar{l}}S^{mk}+\Gamma^{j}_{im}\bar{C}^{mk}_{\bar{l}}\\
&&
+\delta^{k}_{i}\bar{\partial}_{\bar{l}}S^{j}+{\delta^{k}_{n}G_{i\bar{l}}S^{nj}}-C_{int}\bar{C}^{tk}_{\bar{l}}S^{nj}+\Gamma^{k}_{in}\bar{C}^{nj}_{\bar{l}}\quad {-2G_{i\bar{l}}S^{jk}}{-2K_{i}\bar{C}^{jk}_{\bar{l}}}\\
&=&e^{2K}(-\Gamma^{j}_{ir}G^{r\bar{j}})G^{k\bar{k}}\bar{C}_{\bar{j}\bar{k}\bar{l}}+e^{2K}G^{j\bar{j}}
(-\Gamma^{k}_{is}G^{s\bar{k}})\bar{C}_{\bar{j}\bar{k}\bar{l}}\\
&&+\delta^{j}_{i}\overline{\partial}_{\bar{l}}S^{k}-C_{imt}\bar{C}^{tj}_{\bar{l}}S^{mk}+\Gamma^{j}_{im}\bar{C}^{mk}_{\bar{l}}
+\delta^{k}_{i}\bar{\partial}_{\bar{l}}S^{j}-C_{int}\bar{C}^{tk}_{\bar{l}}S^{nj}+\Gamma^{k}_{in}\bar{C}^{nj}_{\bar{l}}\,.
\end{eqnarray*}
According to
$-\bar{C}^{tj}_{\bar{l}}S^{mk}=-\bar{\partial}_{\bar{l}}S^{tj}S^{mk}=-\bar{\partial}_{\bar{l}}(S^{tj}S^{mk})+S^{tj}\bar{\partial}_{\bar{l}}S^{mk}$, we then obtain
\begin{eqnarray*}
&&A+B+C+D\\
&=&{e^{2K}(-\Gamma^{j}_{ir}G^{r\bar{j}})G^{k\bar{k}}\bar{C}_{\bar{j}\bar{k}\bar{l}}}+{e^{2K}G^{j\bar{j}}
(-\Gamma^{k}_{is}G^{s\bar{k}})\bar{C}_{\bar{j}\bar{k}\bar{l}}}\\
&&+\bar{\partial}_{\bar{l}}(\delta^{j}_{i}S^{k}+\delta^{k}_{i}S^{j}-C_{imt}S^{tj}S^{mk})+{C_{imt}S^{tj}\bar{\partial}_{\bar{l}}S^{mk}}{-C_{int}\bar{C}^{tk}_{\bar{l}}S^{nj}}+{\Gamma^{j}_{im}\bar{C}^{mk}_{\bar{l}}}+{\Gamma^{k}_{in}\bar{C}^{nj}_{\bar{l}}}\\
&=&\bar{\partial}_{\bar{l}}(\delta^{j}_{i}S^{k}+\delta^{k}_{i}S^{j}-C_{imt}S^{tj}S^{mk})\,.
\end{eqnarray*}
Therefore,
\begin{equation}\label{DSij}
D_{i}S^{jk}=
\delta^{j}_{i}S^{k}+\delta^{k}_{i}S^{j}-C_{imn}S^{jn}S^{mk}+h_{i}^{jk}
\end{equation}
for some holomorphic quantity $h_{i}^{jk}$.

The quantity $D_{i}S^{j}$ can be obtained in the same manner:
\begin{eqnarray*}
&&\bar{\partial}_{\bar{l}}(D_{i}S^{j})\\
&=&
\bar{\partial}_{\bar{l}}(\partial_{i}S^{j}-2K_{i}S^{j}+\Gamma^{j}_{ir}S^{r})\\
&=&\partial_{i}S^{j}_{\bar{l}}+\bar{\partial}_{\bar{l}}(-2K_{i}S^{j})+\bar{\partial}_{\bar{l}}(\Gamma^{j}_{ir}S^{r})\\
&=&\partial_{i}(G_{l\bar{l}}S^{lj})-2G_{i\bar{l}}S^{j}-2K_{i}\bar{\partial}_{\bar{l}}S^{j}
+\bar{\partial}_{\bar{l}}\Gamma^{j}_{ir}S^{r}+\Gamma^{j}_{ir}\bar{\partial}_{\bar{l}}S^{r}\\
&=&\partial_{i}G_{l\bar{l}}S^{lj}+G_{l\bar{l}}\partial_{i}S^{lj}
-2G_{i\bar{l}}S^{j}-2K_{i}\bar{\partial}_{\bar{l}}S^{j}+\bar{\partial}_{\bar{l}}
\Gamma^{j}_{ir}S^{r}+\Gamma^{j}_{ir}\bar{\partial}_{\bar{l}}S^{r}\\
&=&\partial_{i}G_{l\bar{l}}S^{lj}+G_{l\bar{l}}\partial_{i}S^{lj}
-2G_{i\bar{l}}S^{j}-2K_{i}\bar{\partial}_{\bar{l}}S^{j}+(\delta^{j}_{i}G_{r\bar{l}}+\delta^{j}_{r}G_{i\bar{l}}-C_{irt}\bar{C}^{tj}_{\bar{l}})S^{r}+\Gamma^{j}_{ir}\bar{\partial}_{\bar{l}}S^{r}\\
&=&\partial_{i}G_{l\bar{l}}S^{lj}+G_{l\bar{l}}(D_{i}S^{lj}+{2K_{i}S^{lj}}-\Gamma^{l}_{im}S^{mj}{-\Gamma^{j}_{in}S^{ln})}-2G_{i\bar{l}}S^{j}{-2K_{i}\bar{\partial}_{\bar{l}}S^{j}}\\
&&+(\delta^{j}_{i}
G_{r\bar{l}}+\delta^{j}_{r}G_{i\bar{l}}-C_{irt}\bar{C}^{tj}_{\bar{l}})S^{r}+{\Gamma^{j}_{ir}\bar{\partial}_{\bar{l}}S^{r}}\\
&=&{\Gamma^{m}_{il}G_{m\bar{l}}S^{lj}}+G_{l\bar{l}}(D_{i}S^{lj}{-\Gamma^{l}_{im}S^{mj}})-2G_{i\bar{l}}S^{j}+(\delta^{j}_{i}
G_{r\bar{l}}+\delta^{j}_{r}G_{i\bar{l}}-C_{irt}\bar{C}^{tj}_{\bar{l}})S^{r}\\
&=&G_{l\bar{l}}D_{i}S^{lj}-2G_{i\bar{l}}S^{j}+(\delta^{j}_{i}
G_{r\bar{l}}+\delta^{j}_{r}G_{i\bar{l}}-C_{irt}\bar{C}^{tj}_{\bar{l}})S^{r}\\
&=&G_{l\bar{l}}({\delta^{l}_{i}S^{j}}+
{\delta^{j}_{i}S^{l}}-C_{itm}S^{tl}S^{mj}+h^{lj}_{i})-{2G_{i\bar{l}}S^{j}}+({\delta^{j}_{i}
G_{r\bar{l}}}+{\delta^{j}_{r}G_{i\bar{l}}}-C_{irt}\bar{C}^{tj}_{\bar{l}})S^{r}\\
&=&2\delta^{j}_{i}\bar{\partial}_{\bar{l}}S-G_{l\bar{l}}C_{itm}S^{tl}S^{mj}-C_{irt}\bar{C}^{tj}_{\bar{l}}S^{r}+G_{l\bar{l}}h^{lj}_{i}\,.
\end{eqnarray*}
Since
\begin{eqnarray*}
G_{l\bar{l}}C_{itm}S^{tl}S^{mj}=\bar{\partial}_{\bar{l}}K_{l} C_{imt}S^{tl}S^{mj}=\bar{\partial}_{\bar{l}}S^{t}C_{imt}S^{mj}
=\bar{\partial}_{\bar{l}}(S^{t}C_{imt}S^{mj})-S^{t}C_{imt}\bar{\partial}_{\bar{l}}S^{mj}\,,
\end{eqnarray*}
we then get
\begin{eqnarray*}
\bar{\partial}_{\bar{l}}(D_{i}S^{j})=
2\delta^{j}_{i}\bar{\partial}_{\bar{l}}S-\bar{\partial}_{\bar{l}}(S^{t}C_{imt}S^{mj})+G_{l\bar{l}}h^{lj}_{i}.
\end{eqnarray*}
Hence
\begin{eqnarray}\label{DSi}
D_{i}S^{j}=2\delta^{j}_{i}S-C_{imn}S^{m}S^{nj}+K_{l}h^{lj}_{i}+h^{j}_{i}
\end{eqnarray}
for some holomorphic quantity $h_{i}^{j}$.

In the following we shall calculate $\bar{\partial}_{\bar{l}}D_{i}S$:
\begin{eqnarray*}
&&\bar{\partial}_{\bar{l}}D_{i}S\\
&=& \bar{\partial}_{\bar{l}}(\partial_{i}S-2K_{i}S)\\
&=& \partial_{i}(G_{l\bar{l}}S^{l})+\bar{\partial}_{\bar{l}}(-2K_{i}S)\\
&=& \partial_{i}G_{l\bar{l}}S+G_{l\bar{l}}\partial_{i}S-2\bar{\partial}_{\bar{l}}(K_{i}S)\\
&=& \partial_{i}G_{l\bar{l}}S+G_{l\bar{l}}(D_{i}S^{l}+2K_{i}S-\Gamma^{l}_{im}S^{m})-2\bar{\partial}_{\bar{l}}(K_{i}S)\\
&=&{ G_{m\bar{l}}\Gamma^{m}_{il}S^{l}}+G_{l\bar{l}}({2\delta^{l}_{i}S}-C_{imn}S^{m}S^{nl}+h^{lk}_{i}K_{k}+h^{l}_{i})
+G_{l\bar{l}}({2K_{i}S^{l}}-{\Gamma^{l}_{im}S^{m}})
+{\bar{\partial}_{\bar{l}}(-2K_{i}S)}\\
&=&G_{l\bar{l}}(-C_{imn}S^{m}S^{nl}+h^{lk}_{i}K_{k}+h^{l}_{i})\\
&=&-C_{imn}\bar{\partial}_{\bar{l}}S^{n} S^{m}+\bar{\partial}_{\bar{l}}K_{l} h^{lk}_{i}K_{k}+\bar{\partial}_{\bar{l}}K_{l}h^{l}_{i}\\
&=& -\bar{\partial}_{\bar{l}}({1\over 2}S^{n} C_{imn}S^{m})+\bar{\partial}_{\bar{l}} ({1\over 2}h^{lk}_{i}K_{k}K_{l})+\bar{\partial}_{\bar{l}}(K_{l}h^{l}_{i})\,.
\end{eqnarray*}
So we get
\begin{eqnarray}\label{DS}
D_{i}S= -{1\over 2}C_{imn}S^{m}S^{n} +{1\over 2}h^{kl}_{i}K_{k}K_{l}+K_{l}h^{l}_{i}+h_{i}\,.
\end{eqnarray}
for some holomorphic quantity $h_{i}$.

Now we calculate $D_{i}K_{j}$ as follows. According to
$D_{i}K_{j}=\partial_{i}K_{j}-\Gamma^{m}_{ji}K_{m}$, it follows that
\begin{eqnarray*}
\bar{\partial}_{\bar{l}} (D_{i}K_{j})
&=& \bar{\partial}_{\bar{l}}(\partial_{i}K_{j}-\Gamma^{m}_{ji}K_{m})\\
&=& \partial_{i}K_{j\bar{l}}-\bar{\partial}_{\bar{l}}(\Gamma^{m}_{ji})K_{m}-\Gamma^{m}_{ji}K_{m\bar{l}}\\
&=& {\Gamma^{m}_{ij}G_{m\bar{l}}}-(\delta^{m}_{j}G_{i\bar{l}}+\delta^{m}_{i}G_{j\bar{l}}-C_{jin}\bar{C}^{mn}_{\bar{l}})K_{m}-{\Gamma^{m}_{ji}K_{m\bar{l}}}\\
&=& -G_{i\bar{l}}K_{j}-G_{j\bar{l}}K_{i}+C_{jin}\bar{C}^{mn}_{\bar{l}}K_{m}\\
&=& -\bar{\partial}_{\bar{l}}(K_{i}K_{j})+C_{jin}\bar{\partial}_{\bar{l}}S^{mn} K_{m}\\
&=& -\bar{\partial}_{\bar{l}}(K_{i}K_{j})+\bar{\partial}_{\bar{l}}(C_{jin}S^{mn} K_{m})-C_{jin}S^{mn}\bar{\partial}_{\bar{l}}K_{m}\\
&=& -\bar{\partial}_{\bar{l}}(K_{i}K_{j})+\bar{\partial}_{\bar{l}}(C_{jin}S^{mn} K_{m})-C_{jin}S^{mn}G_{m\bar{l}}\\
&=& -\bar{\partial}_{\bar{l}}(K_{i}K_{j})+\bar{\partial}_{\bar{l}}(C_{jin}S^{mn} K_{m})-C_{jin}\bar{\partial}_{\bar{l}}S^{n}\\
&=& \bar{\partial}_{\bar{l}}(-K_{i}K_{j}+C_{jin}S^{mn} K_{m}-C_{jin}S^{n})\,.
\end{eqnarray*}
Therefore,
\begin{eqnarray}\label{DKi}
D_{i}K_{j}=-K_{i}K_{j}+C_{ijn}S^{mn} K_{m}-C_{ijn}S^{n}+h_{ij}
\end{eqnarray}
for some holomorphic quantity $h_{ij}$.

We want to point out that the holomorphic limit of the differential ring of generators $S^{jk},S^{j},S,K_{i}$ also satisfies similar equations, with everything replaced by their holomorphic limits \cite{Alim:2008kp}.

Some of the holomorphic quantities $h_i^{jk},h_i^j,h_{i},h_{ij}$ can not be uniquely determined \cite{Alim:2008kp}, as briefly discussed in Section \ref{solstogenerators} below, since the above
equations are derived by integrating equations.
Eqs.\,(\ref{DSij}, \ref{DSi}, \ref{DS}, \ref{DKi}) do not
actually give a differential ring due to the existence of these holomorphic
quantities and their derivatives. To make it a genuine ring, one needs to include all of
the derivatives of these holomorphic quantities
\cite{Hosono:2008ve}. In \cite{Yamaguchi:2004bt, Hosono:2008ve, Alim:2013eja},
it is shown that for some special CY 3-fold families, all of the
holomorphic quantities and their derivatives are packaged together by making use of the
special K\"ahler geometry on the moduli space, and are in fact
Laurent polynomials of the Yukawa couplings.
Then one gets a differential ring
with finitely many generators, including the non-holomorphic generators $S^{jk},S^{j},S,K_{i}$ and the holomorphic Yukawa couplings.

Now we shall focus on the cases $h^{2,1}(X)=\dim \mathcal{M}=1$ and consider the differential ring structure of the generators with derivatives taken in the $\tau=(1/ 2\pi i ) \kappa^{-1}F_{tt}$ coordinate, where $\kappa$ is the classical triple intersection (of the A-model CY) and $t$ is the coordinate defined in Eq.\,(\ref{periodsforcompactCYs}). This definition was introduced \cite{Aganagic:2006wq} to match the known modularity for the moduli space of some non-compact CY 3-folds whose geometries are completely determined by the mirror curves sitting inside them, see also \cite{Alim:2013eja, Zhou:2013hpa, Alim:2014gha} for related works.
First one
makes the following change of generators \cite{Alim:2007qj}:
\begin{eqnarray*}
\tilde{S}^{tt} &= & S^{tt},\\
  \tilde{S}^t &= & S^t - S^{tt} K_t,\\
\tilde{S} &= & S- S^t K_t + \frac{1}{2} S^{tt} K_t K_t ,\\
\tilde{K}_t&= &
K_t\,.
\end{eqnarray*}
Then one defines $ \tau=(1/2\pi i)\kappa^{-1}
\partial_{t} F_{t} $ which gives $\frac{\partial \tau}{\partial
t}=(1/2\pi i)\kappa^{-1}C_{ttt}$.
After that one forms the following quantities \cite{Alim:2013eja} on the
deformation space $\mathcal{M}$:
  \begin{eqnarray}\label{dfnofspecialpolynomialgenerators}
     K_0 &= &  \kappa C_{ttt}^{-1} 			(\theta t)^{-3} \,,\nonumber\\
				G_1&= &\theta t\,,\nonumber\\
				 K_2&= &\kappa C_{ttt}^{-1}\tilde{K}_t\,,\nonumber\\
      T_2&= &\tilde{S}^{tt}\,,\nonumber\\
			T_4&= & C_{ttt}^{-1} \tilde{S}^t\,,\nonumber\\
      T_6&= &C_{ttt}^{-2} \tilde{S}\,,
\end{eqnarray}
where $\theta=z{\partial\over \partial z}$ and the propagators $\tilde{S}^{tt},\tilde{S}^{t},\tilde{S}$ are the
normalized (by suitable powers of $X^{0}$ so that they become sections
of $\mathcal{L}^{0}$) propagators. It follows that the
derivatives of the generators given in Eqs.\,(\ref{DSij}, \ref{DSi}, \ref{DS}, \ref{DKi}) now become \cite{Alim:2013eja} the following differential equations satisfied by
$K_{0},G_{1},K_{2},T_{2},T_{4},T_{6}$:
\begin{eqnarray}\label{newformofring}
&\partial_{\tau}K_0&=-2K_0\,K_2- K_0^2\, G_1^2\,(\tilde{h}^z_{zzz}+3(s_{zz}^z+1))\nonumber\,,\\
&\partial_{\tau} G_1&= 2G_1\,K_2-  \kappa G_1\,T_2\,+K_0 G_1^3(s_{zz}^z+1)\nonumber\,,\\
&\partial_{\tau} K_2&=3K_2^2-3 \kappa K_2\,T_2- \kappa^{2}T_4+K_0^2\,G_1^4 k_{zz}-K_0\,G_1^2\,K_2\,\tilde{h}^z_{zzz}\nonumber\,,\\
&\partial_{\tau} T_2&=2K_2\,T_2- \kappa T_2^2+2 \kappa T_4+\kappa^{-1}K_0^2 G_1^4 \tilde{h}^z_{zz}\nonumber\,,\\
&\partial_{\tau} T_4&=4 K_2 T_4-3 \kappa T_2\,T_4+ 2\kappa  T_6-K_0\, G_1^2 \, T_4 \tilde{h}^z_{zzz}- \kappa^{-1}K_0^2\, G_1^4 \,T_2 k_{zz}+\kappa^{-2}K_0^3\, G_1^6 \tilde{h}_{zz}\nonumber\,,\\
&\partial_{\tau} T_6&= 6 K_2\, T_6- 6 \kappa T_2 \,T_6+\frac{\kappa}{2} T_4^2- \kappa^{-1}K_0^2\, G_1^4 \,T_4\,k_{zz}\nonumber\\
&&+\kappa^{-3}K_0^4\, G_1^8 \tilde{h}_z-2 \, K_0\,G_1^2\,T_6
\tilde{h}^z_{zzz}\,,
\end{eqnarray}
where $\partial_{\tau}={1\over 2\pi
i}{\partial\over \partial \tau}$ and the quantities
$\tilde{h}^z_{zzz},s_{zz}^z
,k_{zz},\tilde{h}^z_{zz}
,\tilde{h}_{zz},\tilde{h}_z$ are some holomorphic
functions. It turns out that they are polynomials of the quantity
$C_{0}=\theta \log (z^{3}C_{zzz})$ which satisfies
\begin{equation}
\partial_{\tau} C_{0}=C_{0}(C_{0}+1)G_{1}^{2}\,.
\end{equation}
For special CY 3-fold families these explicit polynomials could be
found in \cite{Alim:2013eja, Alim:2014gha} and will be discussed later in Section \ref{sectionmodularity}.
In the rest of the note, we shall call this particular form of the differential ring the special polynomial ring.

\subsection{Polynomial structure}

Let us resume the discussion on genus two holomorphic anomaly equation.
Direct computation shows that
\begin{eqnarray*}
\widetilde{\mathrm{II}}&=&-{1\over 4}\partial_{k}C_{jmn}S^{jk} \bar{\partial}_{\bar{i}}S^{mn}-{1\over 4}C_{jmn}S^{jk}\bar{\partial}_{\bar{i}}\partial_{k}S^{mn}
-{1\over 2}c_{\chi} S^{jk}\bar{\partial}_{\bar{i}} \partial_{k}K_{j}\\
&=&-{1\over 4}\partial_{k}C_{jmn}S^{jk} \bar{\partial}_{\bar{i}}S^{mn}
-{1\over 4}C_{jmn}S^{jk} \bar{\partial}_{\bar{i}} ( D_{k}S^{mn}+2K_{k}S^{mn}-\Gamma_{kp}^{m}S^{pn}-\Gamma_{kp}^{n}S^{pm})\\
&&-{1\over 2}c_{\chi} S^{jk}\bar{\partial}_{\bar{i}} (D_{k}K_{j}+\Gamma_{kj}^{l}K_{l})\\
&=&-{1\over 4}\partial_{k}C_{jmn}S^{jk} \bar{\partial}_{\bar{i}}S^{mn}\\
&&-{1\over 4}C_{jmn}S^{jk} \bar{\partial}_{\bar{i}} (\delta_{k}^{m}S^{n}+\delta_{k}^{n}S^{m}-C_{kpq}S^{mq}S^{nq})
-{1\over 4}C_{jmn}S^{jk} \bar{\partial}_{\bar{i}} ( 2K_{k}S^{mn}-\Gamma_{kp}^{m}S^{pn}-\Gamma_{kp}^{n}S^{pm})\\
&&
-{1\over 2}c_{\chi} S^{jk}\bar{\partial}_{\bar{i}} (-K_{j}K_{k}+C_{jkp}S^{pq}K_{q}-C_{jkp}S^{p})
-{1\over 2}c_{\chi} S^{jk}\bar{\partial}_{\bar{i}} (\Gamma_{kj}^{l}K_{l})\,.
\end{eqnarray*}
It follows then that
\begin{eqnarray*}
&&\widetilde{\mathrm{I}}+\widetilde{\mathrm{II}}
+\mathrm{III}+\mathrm{IV}\\
&=&
-{1\over 4}C_{jmn}C_{krs}\partial_{\bar{i}} S^{mn}   S^{jk} S^{rs}
-{1\over 2}c_{\chi} C_{jmn}  \partial_{\bar{i}} S^{mn}     S^{jk} K_{k}
-{1\over 2} f_{k}^{(1)} C_{jmn}  \partial_{\bar{i}} S^{mn}     S^{jk}  \\
&&
-{1\over 2} c_{\chi} C_{krs}  \partial_{\bar{i}}K_{j}  S^{kj}  S^{rs}
- c_{\chi}^{2}  \partial_{\bar{i}} (S^{k}K_{k})+ c_{\chi}^{2}  \partial_{\bar{i}} S
-c_{\chi}    \partial_{\bar{i}} (f_{k}^{(1)}S^{k})\\
&+&
-{1\over 4}\partial_{k}C_{jmn}S^{jk} \bar{\partial}_{\bar{i}}S^{mn}\\
&&-{1\over 4}C_{jmn}S^{jk} \bar{\partial}_{\bar{i}}  (\delta_{k}^{m}S^{n}+\delta_{k}^{n}S^{m}-C_{kpq}S^{pm}S^{qn})
-{1\over 4}C_{jmn}S^{jk} \bar{\partial}_{\bar{i}} ( 2K_{k}S^{mn}-\Gamma_{kp}^{m}S^{pn}-\Gamma_{kp}^{n}S^{pm})\\
&&
-{1\over 2}c_{\chi} S^{jk}\bar{\partial}_{\bar{i}} (-K_{j}K_{k}+C_{jkp}S^{pq}K_{q}-C_{jkp}S^{p})
-{1\over 2}c_{\chi} S^{jk}\bar{\partial}_{\bar{i}} \Gamma_{kj}^{l}K_{l}
-{1\over 2}c_{\chi} S^{jk} \Gamma_{kj}^{l}\bar{\partial}_{\bar{i}}K_{l}
\\
&+&
{1\over 4}C_{lmn} S^{jk}\partial_{\bar{i}}  \Gamma^{l}_{jk}S^{mn}
+{1\over 2}c_{\chi}S^{jk}\partial_{\bar{i}}  \Gamma_{jk}^{l} K_{l}
+{1\over 2}f_{l}^{1} S^{jk}\partial_{\bar{i}}  \Gamma_{jk}^{l}\\
&+&
{1\over 4}C_{lmn} S^{jk} \Gamma^{l}_{jk}\partial_{\bar{i}} S^{mn}
+{1\over 2}c_{\chi}S^{jk} \Gamma_{jk}^{l} \partial_{\bar{i}} K_{l}\\
&=&
-{1\over 4}C_{jmn}C_{krs}\partial_{\bar{i}} S^{mn}   S^{jk} S^{rs}
-{1\over 4}\partial_{k}C_{jmn}S^{jk} \bar{\partial}_{\bar{i}}S^{mn}\\
&&
-{1\over 4}C_{jmn}S^{jk} \bar{\partial}_{\bar{i}}  (\delta_{k}^{m}S^{n}+\delta_{k}^{n}S^{m}-C_{kpq}S^{pm}S^{qn})
-{1\over 4}C_{jmn}S^{jk} \bar{\partial}_{\bar{i}} ( 2K_{k}S^{mn}-\Gamma_{kp}^{m}S^{pn}-\Gamma_{kp}^{n}S^{pm}) \\
&&+
{1\over 4}C_{lmn} S^{jk}\partial_{\bar{i}}  \Gamma^{l}_{jk}S^{mn}
+
{1\over 4}C_{lmn} S^{jk} \Gamma^{l}_{jk}\partial_{\bar{i}} S^{mn}\\
&&-{1\over 2} c_{\chi} C_{jkp}  \partial_{\bar{i}} (K_{q} S^{jk} S^{pq})
- {1\over 2}f_{l}^{(1)}C_{jkm}\partial_{\bar{i}}  (S^{jk}S^{ml})\\
&&+f_{j}^{(1)}\partial_{\bar{i}} S^{j}- c_{\chi}^{2}  \partial_{\bar{i}} (S^{k}K_{k})+ c_{\chi}^{2}  \partial_{\bar{i}} S
-c_{\chi}    \partial_{\bar{i}} (f_{k}^{(1)}S^{k})+c_{\chi}\partial_{\bar{i}} (S^{j}K_{j})-c_{\chi} \partial_{\bar{i}} S\,.
\end{eqnarray*}
Now we make use of the fact that $D_{k}C_{jmn}=D_{k}D_{j}D_{m}D_{n}\mathcal{F}$ is symmetric in $j,k,m,n$ to simplify the above expression.
By definition, we have
\begin{equation}
\partial_{k}C_{jmn}=D_{k}C_{jmn}-2K_{k}C_{jmn}+\Gamma_{kj}^{l}C_{lmn}
+\Gamma_{km}^{l}C_{jln}+\Gamma_{kn}^{l}C_{jml}\,.
\end{equation}
Then
\begin{eqnarray*}
&&\partial_{k}C_{jmn}S^{jk}\bar{\partial}_{\bar{i}}S^{mn}\\
&=&D_{k}C_{jmn}S^{jk}\bar{\partial}_{\bar{i}}S^{mn}-2K_{k}C_{jmn}S^{jk}\bar{\partial}_{\bar{i}}S^{mn}\\
&&+\Gamma_{kj}^{l}C_{lmn}S^{jk}\bar{\partial}_{\bar{i}}S^{mn}
+\Gamma_{km}^{l}C_{jln}S^{jk}\bar{\partial}_{\bar{i}}S^{mn}+\Gamma_{kn}^{l}C_{jml}S^{jk}\bar{\partial}_{\bar{i}}S^{mn}\\
&=&{1\over 2}(\bar{\partial}_{\bar{i}}(D_{k}C_{jmn}S^{jk}S^{mn})-\bar{\partial}_{\bar{i}}D_{k}C_{jmn} S^{jk}S^{mn})\\
&&-2K_{k}C_{jmn}S^{jk}\bar{\partial}_{\bar{i}}S^{mn}+\Gamma_{kj}^{l}C_{lmn}S^{jk}\bar{\partial}_{\bar{i}}S^{mn}
+\Gamma_{km}^{l}C_{jln}S^{jk}\bar{\partial}_{\bar{i}}S^{mn}+\Gamma_{kn}^{l}C_{jml}S^{jk}\bar{\partial}_{\bar{i}}S^{mn}\\
&=&
{1\over 2}\bar{\partial}_{\bar{i}}(D_{k}C_{jmn}S^{jk}S^{mn})\\
&&-{1\over 2} (2C_{jmn}\bar{\partial}_{\bar{i}} K_{k}S^{jk}S^{mn}
-C_{lmn}\bar{\partial}_{\bar{i}}\Gamma^{l}_{kj}S^{jk}S^{mn}
-C_{jln}\bar{\partial}_{\bar{i}}\Gamma^{l}_{km}S^{jk}S^{mn}
-C_{jml}\bar{\partial}_{\bar{i}}\Gamma^{l}_{kn}S^{jk}S^{mn})\\
&&-2K_{k}C_{jmn}S^{jk}\bar{\partial}_{\bar{i}}S^{mn}+\Gamma_{kj}^{l}C_{lmn}S^{jk}\bar{\partial}_{\bar{i}}S^{mn}
+\Gamma_{km}^{l}C_{jln}S^{jk}\bar{\partial}_{\bar{i}}S^{mn}+\Gamma_{kn}^{l}C_{jml}S^{jk}\bar{\partial}_{\bar{i}}S^{mn}\,.
\end{eqnarray*}
Therefore, we obtain
\begin{eqnarray*}
&&\widetilde{\mathrm{I}}+\widetilde{\mathrm{II}}
+\mathrm{III}+\mathrm{IV}\\
&=& -{1\over 8}(\bar{\partial}_{\bar{i}} D_{k}C_{jmn})
S^{jk}S^{mn}
+{1\over 4}C_{jmn}S^{jk}C_{kpq} S^{pm}\bar{\partial}_{\bar{i}}S^{qn}\\
&&-{1\over 4}C_{jpq}C_{krs}\partial_{\bar{i}} S^{pq}   S^{jk} S^{rs}
-{1\over 8}C_{lmn}C_{jkp}S^{jk}\partial_{\bar{i}} S^{pl}   S^{mn} \\
&&-{1\over 2} c_{\chi} C_{jkp}  \partial_{\bar{i}} (K_{q} S^{jk} S^{pq})
- {1\over 2}f_{l}^{(1)}C_{jkp}\partial_{\bar{i}}  (S^{jk}S^{pl})\\
&&+f_{j}^{(1)}\partial_{\bar{i}} S^{j}- c_{\chi}^{2}  \partial_{\bar{i}} (S^{k}K_{k})+ c_{\chi}^{2}  \partial_{\bar{i}} S
-c_{\chi}    \partial_{\bar{i}} (f_{k}^{(1)}S^{k})+c_{\chi}\partial_{\bar{i}} (S^{j}K_{j})-c_{\chi} \partial_{\bar{i}} S\\
&=& -{1\over 8}(\bar{\partial}_{\bar{i}} D_{k}C_{jmn})
S^{jk}S^{mn}
+{1\over 12}C_{jmn}C_{kpq} \bar{\partial}_{\bar{i}}(S^{jk}S^{pm}S^{qn})-{1\over 8}C_{jpq}C_{krs}\partial_{\bar{i}} (S^{pq}   S^{jk} S^{rs})
 \\
&&-{1\over 2} c_{\chi} C_{jkp}  \partial_{\bar{i}} (K_{q} S^{jk} S^{pq})
- {1\over 2}f_{l}^{(1)}C_{jkp}\partial_{\bar{i}}  (S^{jk}S^{pl})\\
&&+f_{j}^{(1)}\partial_{\bar{i}} S^{j}- c_{\chi}^{2}  \partial_{\bar{i}} (S^{k}K_{k})+ c_{\chi}^{2}  \partial_{\bar{i}} S
-c_{\chi}    \partial_{\bar{i}} (f_{k}^{(1)}S^{k})+c_{\chi}\partial_{\bar{i}} (S^{j}K_{j})-c_{\chi} \partial_{\bar{i}} S\,.
\end{eqnarray*}
It follows that
\begin{eqnarray}\label{DF2}
&& \partial_{\bar{i}}\mathcal{F}^{(2)}\\
&=&
 \partial_{\bar{i}}[
-{1\over 8}(D_{k}C_{jmn})
S^{jk}S^{mn}
+{1\over 12}C_{jmn}C_{kpq} S^{jk}S^{pm}S^{qn}-{1\over 8}C_{jpq}C_{krs} S^{pq}   S^{jk} S^{rs}\nonumber\\
&&
-{1\over 2} c_{\chi} C_{jkp} K_{q} S^{jk} S^{pq}
- {1\over 2}f_{l}^{(1)}C_{jkp}  S^{jk}S^{pl}\nonumber\\
&&+f_{j}^{(1)} S^{j}- c_{\chi}^{2}   S^{k}K_{k}+ c_{\chi}^{2}   S
-c_{\chi}    f_{k}^{(1)}S^{k}+c_{\chi}S^{j}K_{j}-c_{\chi} S]\nonumber\,.
\end{eqnarray}
Therefore, we can see up to the addition by a holomorphic ambiguity $f^{(2)}$, the genus two topological string partition function $\mathcal{F}^{(2)}$ is a polynomial of the propagators $S^{jk},S^{j},S$ and the generators $K_{i}$.

In fact, it was originally shown in \cite{Bershadsky:1993cx} that a solution of the recursion holomorphic anomaly equations is given in terms of Feynman
rules. The propagators $S^{ij}$, $S^i$, $S$
for these Feynman rules were defined in Eqs.~(\ref{defofSij},\ref{defofSi}, \ref{defofS}):
\begin{equation}\label{defofpropagators}
 \partial_{\bar{i}}S^{jk}=\overline{C}_{\bar{i}}^{jk},\quad
 \partial_{\bar{i}}S^{j}=G_{\bar{i}k}S^{jk},\quad
 \partial_{\bar{i}}S=G_{\bar{i}k}S^{k}.
\end{equation}
The vertices of the Feynman rules are
given by the functions $\mathcal{F}^{(g)}_{i_1\cdots
i_n}=D_{i_{1}}\cdots D_{i_{n}}\mathcal{F}^{(g)}$.
For example, for genus two, the above topological string partition function in Eq.\,(\ref{DF2}) has the form
\begin{eqnarray*}
\mathcal{F}^{(2)}&=&
{1\over 2}S^{jk}D_{j}D_{k}\mathcal{F}^{(1)}
+{1\over 2}S^{jk}D_{j}\mathcal{F}^{(1)}
D_{k}\mathcal{F}^{(1)}
-{1\over 8}S^{jk}S^{mn}D_{j}D_{k}D_{m}D_{n}\mathcal{F}\\
&&-{1\over 2}S^{jk}C_{jkm}S^{mn}D_{n}\mathcal{F}^{(1)}
{\chi\over 24}S^{j}D_{j}\mathcal{F}^{(1)}
+
{1\over 8}S^{jk}C_{jkp}S^{pq}C_{qmn}S^{mn}\\
&&
+{1\over 12}S^{jk}S^{pq}S^{mn}C_{jpm}C_{kqn}
-{\chi\over 48}S^{j}C_{jkl}S^{kl}
+{\chi\over 24}({\chi\over 24}-1)S+f^{(2)}\,,
\end{eqnarray*}
for some holomorphic function $f^{(2)}$.
Each of the terms involving the propagators $S^{jk},S^{j},S$
has a diagrammatic interpretation and corresponds to the contribution
of a specific boundary component of the moduli space of
genus two stable curves, see \cite{Bershadsky:1993cx} for details.

Motivated by \cite{Bershadsky:1993cx}, in \cite{Yamaguchi:2004bt, Alim:2007qj} it was proved, using Eqs.\,(\ref{specialgeometryrelation}, \ref{integratedspecialgeometryrelation}, \ref{soltogenusone}, \ref{integratedgenusone}),
that the holomorphic anomaly equations Eq.\,(\ref{haeforhighergenus}) for $g\geq 2$ can be put into
the following form
\begin{equation}\label{polynomialsol}
\bar{\partial}_{\bar{i}}\mathcal{F}^{(g)}=\bar{\partial}_{\bar{i}}\mathcal{P}^{(g)}\,,
\end{equation}
where $\mathcal{P}^{(g)}$ is a polynomial in the generators
$S^{jk},S^{j},S,K_{i}$ with the coefficients being holomorphic
quantities which might have poles. The proof relies on the fact that these generators form a differential ring \cite{Alim:2007qj} as displayed in Eqs.\,(\ref{DSij}, \ref{DSi}, \ref{DS}, \ref{DKi}) and recalled below:
\begin{eqnarray}\label{polyring}
D_i S^{jk} &=&\delta_i^j S^k+\delta_{i}^{k}S^{j}
-C_{imn} S^{mj} S^{nk}  + h_i^{jk} \, , \nonumber\\
D_i S^j &=& -C_{imn}S^{m}S^{jn}+2\delta^j_i S +h_{i}^{jk}K_{k}+h_i^j\,,\nonumber\\
D_i S&=& -\frac{1}{2} C_{imn} S^m S^n+{1\over 2}h_{i}^{mn}K_{m}K_{n}+h_i^{m}K_{m}+h_{i} \, ,\nonumber\\
D_i K_j &=& -K_i K_j +C_{ijm} S^{mn}K_{n} -C_{ijm}S^{m}+ h_{ij} \, ,
\end{eqnarray}
where $h_i^{jk},h_i^j,h_{i},h_{ij}$ are holomorphic quantities.

Now we justify the structure in Eq.\,(\ref{polynomialsol}) by induction, following \cite{Alim:2007qj}.
Note that the non-holomorphicity of the topological string
partition functions only comes from the non-holomorphic generators $S^{ij},S^{i},S,K_{i}$ and thus the
anti-holomorphic derivative on the left-hand side of the holomorphic
anomaly equations can be replaced by derivatives with respect to
these generators.
Furthermore,
one can make a change of generators \cite{Alim:2007qj}:
\begin{eqnarray}\label{shiftofgenerators}
\tilde{S}^{ij} &=& S^{ij}, \nonumber \\
\tilde{S}^i &=& S^i - S^{ij} K_j, \nonumber \\
\tilde{S} &=& S- S^i K_i + \frac{1}{2} S^{ij} K_i K_j, \nonumber\\
\tilde{K}_i&=& K_i\, .
\end{eqnarray}
The differential ring structure among these new non-holomorphic generators follows from Eq.\,(\ref{polyring}) easily.
Replacing the $\bar{\partial}_{\bar{i}}$ derivative in the holomorphic anomaly equations by derivatives with respect to the new non-holomorphic generators and using the definitions Eq.\,(\ref{defofpropagators}), one then gets,
\begin{eqnarray*}\label{polrec1}
&&\bar{\partial}_{\bar{i}} \mathcal{F}^{(g)} \\
&=&  \bar{C}_{\bar{i}}^{jk} \left(  \frac{\partial \mathcal{F}^{(g)}}{\partial S^{jk}} -\frac{1}{2}  \frac{\partial \mathcal{F}^{(g)}}{\partial \tilde{S}^{k}}\tilde{K}_j -\frac{1}{2}  \frac{\partial \mathcal{F}^{(g)}}{\partial \tilde{S}^{j}}\tilde{K}_k + \frac{1}{2}  \frac{\partial \mathcal{F}^{(g)}}{\partial \tilde{S}}\tilde{K}_j \tilde{K}_k\right)+G_{\bar{i}j} \frac{\partial \mathcal{F}^{(g)}}{\partial \tilde{K}_{j}} \\
&=& \frac{1}{2} \bar{C}_{\bar{i}}^{jk} \left(
\sum_{r=1}^{g-1}
D_j\mathcal{F}^{(r)} D_k\mathcal{F}^{(g-r)} +
D_j D_k\mathcal{F}^{(g-1)} \right)\,.
\end{eqnarray*}
Assuming the independence\footnote{This assumption is reasonable since these quantities have different singular behaviors
when written in the canonical coordinates at the large complex structure.} of $\bar{C}_{\bar{i}}^{jk}$ and $G_{\bar{i}j}$, then one gets two sets of equations:
\begin{eqnarray}
 \frac{\partial \mathcal{F}^{(g)}}{\partial \tilde{S}^{jk}} &-&\frac{1}{2} \frac{\partial \mathcal{F}^{(g)}}{\partial \tilde{S}^{k}}\tilde{K}_j -\frac{1}{2}  \frac{\partial \mathcal{F}^{(g)}}{\partial \tilde{S}^{j}}\tilde{K}_k + \frac{1}{2}  \frac{\partial \mathcal{F}^{(g)}}{\partial \tilde{S}}\tilde{K}_j \tilde{K}_k\nonumber\\
&=&
\sum_{r=1}^{g-1}
D_j\mathcal{F}^{(r)} D_k\mathcal{F}^{(g-r)} +
D_jD_k\mathcal{F}^{(g-1)} \nonumber\label{firstsetofeqns}\,,\\
\frac{\partial \mathcal{F}^{(g)}}{\partial \tilde{K}_{j}} &=&0\label{secondsetofeqns}\,.
\end{eqnarray}
Eq.\,(\ref{polynomialsol}) then follows from the above equations
and Eq.\,(\ref{polyring}).

The polynomial structure given in Eq.\,(\ref{polynomialsol}) also
allows to determine the non-holomorphic part $\mathcal{P}^{(g)}$ of $\mathcal{F}^{(g)}$ genus by genus recursively from Eq.\,(\ref{firstsetofeqns})
as polynomials of the new non-holomorphic generators $\tilde{S}^{ij},\tilde{S}^{i},\tilde{S},\tilde{K}_{i}$ and thus of the old ones $S^{ij},S^{i},S,K_{i}$:
\begin{equation}\label{polyrecursion}
\mathcal{P}^{(g)}=\mathcal{P}^{(g)}(S^{ij},S^{i},S,K_{i})\,.
\end{equation}
Moreover, the coefficients of the monomials in these non-holomorphic generators are explicit Laurent polynomials in the holomorphic generators, with
the coefficients of the monomials in the non-holomorphic and holomorphic generators being universal constants. These constants come from the Feynman diagram interpretation \cite{Bershadsky:1993cx}, or equivalently, the combinatorics from recursion. They are independent of the geometry under consideration.
For example, for any geometry, the highest power of $S^{ij}$ in the genus two partition function in Eq.\,(\ref{DF2})
always takes the form ${1\over 12}C_{ipm}C_{jqn}S^{ij}S^{pq}S^{mn}+\cdots$.

\subsection{Solutions of propagators}\label{solstogenerators}

To obtain explicit results for $\mathcal{P}^{(g)}$,
one needs to get formulas for the propagators $S^{ij},S^{i},S, K_{i}$.
The generator $K_{i}$ could be obtained by using its definition as the K\"ahler potential for the Weil-Petersson geometry
Eq.\,(\ref{Kahlerpotential}) and the periods of the CY family Eq.\,(\ref{periodsforcompactCYs}).
The generators $S^{ij},S^{i},S$ can be solved from Eq.\,(\ref{defofpropagators}), up to addition by holomorphic quantities. A special set of solutions whose holomorphic limits are vanishing was given by \cite{Alim:2008kp, Hosono:2008ve} in terms of geometric quantities.
Alternatively, they could be determined \cite{Alim:2008kp} from the differential structure
Eq.\,(\ref{polyring}):
\begin{eqnarray}\label{sol1ofpropagators}
S^{kl}&=&(C_{*}^{-1})^{kj} (\delta^{l}_{*}K_{j}+\delta^{l}_{j}K_{*}-\Gamma^{l}_{*j}+s^{l}_{*j})\nonumber\,,\\
S^{i}&=&{1\over 2} (D_{i}S^{ii}+C_{imn}S^{mi}S^{ni}-h^{ii}_{i})\nonumber\,,\\
S&=&{1\over 2}(D_{i}S^{i}+C_{imn}S^{m}S^{ni}-h^{ik}_{i}K_{k}-h^{i}_{i})\,,
\end{eqnarray}
or equivalently
\begin{eqnarray}\label{sol2ofpropagators}
S^{kl}&=&(C_{*}^{-1})^{kj} (\delta^{l}_{*}K_{j}+\delta^{l}_{j}K_{*}-\Gamma^{l}_{*j}+s^{l}_{*j})\nonumber\,,\\
S^{i}&=&(C_{*}^{-1})^{ij}  (-D_{*}K_{j}-K_{*}K_{j}+C_{*jk}S^{kl}K_{l}+h_{*j})\nonumber\\
&=&(C_{*}^{-1})^{ij}(K_{*}K_{j}-K_{*j}+h^{l}_{*j}K_{l}+h_{*j})\nonumber\,,\\
S&=&{1\over 2}(D_{i}S^{i}+C_{imn}S^{m}S^{ni}-h^{ik}_{i}K_{k}-h^{i}_{i})\,,
\end{eqnarray}
where the sub-index $*$ is chosen so that the matrix $((C_{*})_{ij})$ is invertible. These formulas are useful \cite{Alim:2008kp} in analyzing the degrees of freedom of the holomorphic quantities $h_{i}^{jk},h_{i}^{j},h_{i},h_{ij}$. Moreover, in the one-modulus case for which the dimension of $\mathcal{M}$ is one, they also tell that the set of non-holomorphic generators $S^{zz},S^{z},S,K_{z}$ is equivalent to the set of generators
$\Gamma_{zz}^{z}, K_{zzz}, K_{zz},K_{z}$ considered in \cite{Yamaguchi:2004bt} and the differential rings are thus identical.
By using the latter set of generators, the differential structure
follows from the Picard-Fuchs equation for the quantity $e^{-K}$ and the special geometry relation Eq.\,(\ref{specialgeometryrelation}), see \cite{Yamaguchi:2004bt} for details.

For non-compact CY 3-folds, one can choose a very simple set of generators due to the existence of a constant period.
More precisely, according to Eqs.\,(\ref{Kahlerpotential}, \ref{periodsforcompactCYs}), one can see that the holomorphic limit of $K_{i}$ is zero. Then from Eq.\,(\ref{sol2ofpropagators}) it follows that by choosing vanishing $h_{ij},h_{i}^{i}$ we can arrange so that the holomorphic limits of $S^{i},S$ are zero, see \cite{Alim:2008kp} for more detailed discussions on this. This sometimes makes the calculations easier when computing the quantity $\mathcal{P}^{(g)}$ from recursion.

\section{Connection to modular forms}\label{sectionmodularity}

After determining $\mathcal{P}^{(g)}$ from recursion by using the polynomial structure, one can then try to fix the holomorphic ambiguity $f^{(g)}$ by using the boundary conditions.
We are not going to explain this due to limit of space, but we refer the interested reader to e.g., \cite{Bershadsky:1993ta, Bershadsky:1993cx}.\footnote{See also \cite{Marino:1998pg, Katz:1999xq, Klemm:1999gm, Klemm:2004km, Yamaguchi:2004bt, Klemm:2005pd, Huang:2006hq, Aganagic:2006wq, Huang:2006si, Alim:2007qj, Grimm:2007tm, Alim:2008kp, Haghighat:2008gw, Haghighat:2009nr, Sakai:2011xg, Alim:2012ss, Klemm:2012sx, Alim:2013eja} for related works.}

In this section we shall emphasize the connection (\cite{Huang:2006si, Aganagic:2006wq, Hosono:2008ve, Alim:2013eja, Zhou:2013hpa, Alim:2014gha})
between the differential ring generated by $S^{ij},S^{i},S,K_{i}$ and the differential ring of modular forms. The reason for this expectation is that in some nicest cases, the topological string partition functions are known to be almost-holomorphic modular forms (see for example \cite{Dijkgraaf:1995, Kaneko:1995, Milanov:2011ku, Li:2011mi}) and are thus polynomials in the generators of the ring of almost-holomorphic modular forms. The similarity between the differential ring structure in the propagators/holomorphic limit of propagators of the special polynomial ring Eq.\,(\ref{newformofring}) and the polynomial structure of the ring of almost-holomorphic/quasi-modular forms seems to suggest a connection between them.\\

For the $K_{\mathbb{P}^{2}}$ family (each member of the family is the total space of $K_{\mathbb{P}^{2}}=\mathcal{O}_{\mathbb{P}^{2}}(-3)$, but the K\"ahler structure varies),
the mirror family can be constructed following the lines in \cite{Chiang:1999tz} using Batyrev toric duality \cite{Batyrev:1994hm}, or using the Hori-Vafa construction \cite{Hori:2000kt}.
For definiteness, the equation for the mirror family $\mathcal{X}\to \mathcal{M}$ obtained by the Hori-Vafa method is displayed below:
\begin{equation*}
uv-H(y_{1},y_{2};z)=0,\quad  (u,v,y_{1},y_{2})\in \mathbb{C}^{2}\times (\mathbb{C}^{*})^{2}\,,
\end{equation*}
where $H(y_{1},y_{2};z)=y_{1}y_{2}(-z+y_{1}+y_{2})+1$ and $z$ is the parameter for the base $\mathcal{M}$.
The mirror family $\mathcal{X}\to \mathcal{M}$ comes with the following Picard-Fuchs equation:
\begin{equation*}
\left(\theta^{3}-27z\theta (\theta+{1\over 3})(\theta+{2\over 3})\right)\phi=0~ \textrm{for~a~period } \phi\,,\theta=z{\partial \over \partial z}\,.
\end{equation*}
The coordinate $z$ on $\mathcal{M}$ is chosen so that $z=0$ is the large complex structure limit and $\Delta=1-27z$ is the discriminant.
For this case, the quantity $\kappa$ which appears in Eq.~(\ref{newformofring}) is $-{1\over 3}$.
Then in Eq.~(\ref{dfnofspecialpolynomialgenerators}) we have\footnote{This is due to properties of special K\"ahler geometry
and the particular form for the Picard-Fuchs equation, see \cite{Bershadsky:1993cx} for details.} $\tilde{K}_{t}=0$, we can arrange so that
\begin{equation*}
T_{4}=T_{6}=0\,,
\end{equation*}
by choosing
\begin{equation*}
s_{zz}^z=-\frac{4}{3}+\frac{1}{6\Delta}\,,\quad \tilde{h}^z_{zz}=\frac{1}{36\Delta^2}\,,\quad \tilde{h}^z_{zzz}=\frac{1}{2\Delta}\,.
\end{equation*}
It follows that the special polynomial ring in Eq.\,(\ref{newformofring})
becomes \cite{Alim:2013eja}:
\begin{eqnarray*}
  \partial_{\tau} C_0&=& G_1^2\,C_0\,,\\
  \partial_{\tau} G_1&=&\frac{1}{6}\left(2G_1\, T_2+G_1^3\left(\frac{C_0-1}{C_0+1}\right)\right)\,,\\
 \partial_{\tau} T_2&=&\frac{T_2^2}{3}-\frac{G_1^4}{12}\,.
\end{eqnarray*}
Moreover, in \cite{Alim:2013eja} it was shown that for
this case the moduli space $\mathcal{M}$ can be identified with the modular curve $\Gamma_{0}(3)\backslash \mathcal{H}^{*}$. Under this identification, the points $z=0,1/27$ are identified with the cusp classes $[\tau]=[i\infty],[0]$ respectively. The generators given in Eq.~(\ref{dfnofspecialpolynomialgenerators}) which
can be computed by using Eq.\,(\ref{sol1ofpropagators}) or Eq.\,(\ref{sol2ofpropagators}) are as follows:
\begin{eqnarray*}
T_2&=&{1\over 8}(E_{2}(\tau)+3E_{2}(3\tau))\nonumber\,,\\
G_{1}&=&
\theta_{2}(2\tau)\theta_{2}(6\tau)+\theta_{3}(2\tau)\theta_{3}(6\tau)\nonumber\,,\\
C_0&=&27 \left({\eta(3\tau)\over \eta(\tau)}\right)^{12}.
\end{eqnarray*}
They are essentially the generators for the ring \cite{Kaneko:1995}  of quasi-modular forms $\widehat{M}(\Gamma_{0}(3),\chi_{-3})$ for the modular group $\Gamma_{0}(3)$, with the sub-indices corresponding to the modular weights. The differential ring structure in Eq.~(\ref{newformofring})
corresponds to the Ramanujan-like identities among these quasi-modular forms, see \cite{Aganagic:2006wq, Zagier:2008, Maier:2009, Alim:2013eja, Zhou:2013hpa} and references therein for more details on this ring and its application in studying the modularity of topological string partition functions.

Similarly, for certain mirror families of $K_{\mathrm{dP}_{n}}, n=5,6,7,8$, one can identify \cite{Alim:2013eja} the moduli spaces with certain modular curves and
show that the differential rings Eq.\,(\ref{newformofring})
are identical to the differential rings of quasi-modular forms
with respect to corresponding modular groups.

Assuming the validity of mirror symmetry for these families of CY 3-folds, we can then show that if the solutions to the holomorphic anomaly equations for the B-model CY 3-fold family exist and are unique, then the generating functions of Gromov-Witten invariants for the A-model CY 3-fold family are quasi-modular forms \cite{Alim:2013eja}.
In fact, the existence and uniqueness for the $K_{\mathbb{P}^{2}}$ case are proved in \cite{Zhou:2014thesis} which imply a version of integrality for the sequence of Gromov-Witten invariants $\{N_{g,d}\}_{d=1}^{\infty}$ for any (fixed) $g$. \\

For the mirror quintic family\cite{Candelas:1990rm} $\pi: \mathcal{X}\rightarrow \mathcal{M}$,
the Picard-Fuchs equation is given by
\begin{equation*}
\left(\theta^{4}-5^{5}z (\theta+{1\over 5})(\theta+{2\over 5}) (\theta+{3\over 5}) (\theta+{4\over 5})\right)\phi=0~ \textrm{for~a~period } \phi,\, \theta=z{\partial \over \partial z}\,.
\end{equation*}
We choose the usual $z$ coordinate as the local coordinate on the moduli space $\mathcal{M}$ so that
\footnote{This is related to the $\psi$ coordinate in \cite{Candelas:1990rm} by $z=(5\psi)^{-5}$.} $z=0$
gives the large complex structure limit and the discriminant is $\Delta=1-5^{5}z$.
In this case, the classical triple intersection in Eq.\,(\ref{newformofring}) is $\kappa=5$.
We then choose the following ambiguities \cite{Yamaguchi:2004bt, Hosono:2008ve, Alim:2012gq, Alim:2013eja}:
\begin{equation*}
s_{zz}^{z}=-{8\over 5},
\tilde{h}_{zz}^{z}={1\over 5\Delta},
 \tilde{h}_{zz}=-{1\over 5^{2}\Delta},
\tilde{h}_{z}={2\over 5^{3}\Delta},
k_{zz}={2\over 5^{2}}\,.
\end{equation*}
Now the holomorphic quantities in the ring
Eq.\,(\ref{newformofring}) are polynomials in ${1\over \Delta}$.
Moreover, we also have $C_{0}={5^{5}z\over 1-5^{5}z}={1\over \Delta}-1$ with
$\partial_{\tau}C_{0}=C_{0}(C_{0}+1)K_{0}G_{1}^{2}$
which tells that the derivative of ${1\over \Delta}$ sits inside the ring generated by $K_{0},G_{1},K_{2},T_{2},T_{4},T_{6},C_{0}$.
More examples on compact CY geometries can be found in \cite{Alim:2014gha}.

For compact CY 3-folds, the period domain is in general not Hermitian symmetric and the suitable theory of almost-holomorphic modular forms and quasi-modular forms is not known. Hence one can not say much about the connection between this ring and the ring of modular objects. Nevertheless, the rings of quantities generated by $K_{0},G_{1},K_{2},T_{2},T_{4},T_{6},C_{0}$, which are defined from the special K\"ahler geometry on the moduli spaces of complex structures of CY 3-folds, share very similar properties to those of the rings of almost-holomorphic modular forms defined on modular curves, see \cite{Hosono:2008ve, Zhou:2013hpa} for more discussions on this.

\section{Conclusions}\label{sectionconclusion}

We first showed how to solve for the non-holomorphic part of $\mathcal{F}^{(2)}$ by introducing the propagators $S^{ij},S^{i}, S$.
We then derived the differential ring structure of the ring generated by these propagators and $K_{i}$. After that we proved by induction that
for any $g$, the non-holomorphic part $\mathcal{P}^{(g)}$ is a polynomial of the generators and can be solved recursively genus by genus.
For some special non-compact CY 3-fold families, we pointed out that after the identification between the moduli spaces of complex structures with modular curves, the generators for the special polynomial ring become the generators for the ring of quasi-modular forms, and the differential ring structure is identified with the Ramanujan-like identities for the quasi-modular forms.

It would be interesting to see whether these rings could help construct ring of modular objects (see for example \cite{Movasati:2011zz}), and how the global properties of the generators could help solve for the topological string partition functions from the holomorphic anomaly equations with boundary conditions for more general CY 3-fold families.


%
%
%


\newcommand{\etalchar}[1]{$^{#1}$}
\providecommand{\bysame}{\leavevmode\hbox to3em{\hrulefill}\thinspace}
\providecommand{\MR}{\relax\ifhmode\unskip\space\fi MR }
\providecommand{\MRhref}[2]{%
  \href{http://www.ams.org/mathscinet-getitem?mr=#1}{#2}
}
\providecommand{\href}[2]{#2}

\end{document}